\documentclass[preprint,aps,12pt]{revtex4}
\usepackage{setspace}
\usepackage{amsmath}
\usepackage{graphicx}
\usepackage{graphics}

\begin{document}

\title{Radiation from charges in the continuum limit}
\author{Reuven Ianconescu}
\email{riancon@mail.shenkar.ac.il, riancon@gmail.com}
\affiliation{Shenkar College of Engineering and Design}

\date{\today}

\begin{abstract}
It is known that an accelerating charge radiates according to Larmor
formula. On the other hand, any DC current following a curvilinear
path, e.g. a circular loop, consists of accelerating charges, but in
such case the radiated power is 0. The scope of this paper is to
analyze and quantify how the radiation vanishes when one goes to the
continuum DC limit.
\end{abstract}


\maketitle

\section{Introduction}

There are many physical configurations in which discreet charges are
in acceleration, but in spite of that they behave almost like steady
state DC. This situation is encountered in any DC or low frequency
electrical circuit, because in spite of the fact that the charge
distribution is always discreet and each charge accelerates in the
influence of the electric field, one may consider the charge
distribution as almost continuous. For example the model used for
conducting materials is of an average drift velocity for positive and negative
charges which is proportional to the electric field
$\mathbf{v}^{\pm}=\pm\mu^{\pm} \mathbf{E}$,
$\mu^{\pm}$ being the mobility of the positive and negative charges,
respectively. Hence the current density is
$\mathbf{J}=\rho^{+}\mathbf{v}^{+}+\rho^{-}\mathbf{v}^{-}=(\rho^{+}\mu^{+}-\rho^{-}\mu^{-}) \mathbf{E}$, where
$\rho^{\pm}$ is the charge density of the positive and negative charge
carriers, $\rho^{-}$ being negative, so that the total charge density
is $\rho=\rho^{+}+\rho^{-}=\epsilon_0\mathbf{\nabla}\cdot\mathbf{E}$.
In case the charges are protons and electrons,
in a conductor, the protons mobility is $\mu^{+}=0$ and $\rho^{+}=-\rho^{-}$,
so that usually $\mathbf{\nabla}\cdot\mathbf{E}=0$ like in free space.
In addition, the model considers $\rho^{-}$ to be almost uniform,
so that one defines the conductivity $\sigma=-\rho^{-}\mu^{-}$,
resulting in Ohm's law $\mathbf{J}=\sigma \mathbf{E}$.

Another situation, for which charges are somehow ``more discreet'',
are DC or low frequency ion drift devices, and for those we usually
have one type of charge carriers, say positive ions. Here we use
$\mathbf{J}=\rho \mathbf{v}=\rho\mu \mathbf{E}$, and one may not assume the charge density is uniform,
but rather has to use Gauss's law $\epsilon_0\nabla\cdot\mathbf{E}=\rho$,
resulting in a set of nonlinear equations. In principle, such problem
is time dependent, because discreet ions move in the space, but it
appears that the DC approximation $\nabla\cdot\mathbf{J}=0$ works
very well for such cases \protect\cite{ianc_so_mud,zhao-adamiak,sigmond,sigmond1}.

In either of the above situations, currents may follow curvilinear
paths, in which case, the charges clearly accelerate, also if
the magnitude of their velocity is constant, and still the DC
model works well for most practical cases for which the
density of the charge carriers is big.

The purpose of this work is to understand how the radiation vanishes
when the density of the charge carriers approaches the continuum
steady state. Certainly, the radiation has to disappear gradually, so
that this vanishing may be quantified. Some preliminary work
\protect\cite{plasma_rad} has
been done in this direction, but because this work has been done a priori
in a non relativistic approach, its results are inaccurate and incomplete.

The calculation is done in a canonical configuration of charges in
circular motion at constant speed. The configuration and the formulation
are explained in section~2. In section~3 we calculate the fields, explain their
behavior and derive an exact expression for the radiated power. For completeness,
we also calculate the radiation reaction and show that the power needed to
support the radiation equals the radiated power. In section~4
we derive an asymptotic result for the case the number of charges is big 
(continuum limit). As a special case, we also derive the limit for
slow charges, and this case is of importance because it represents
the typical situation of DC currents in devices as discussed before.
The paper is ended with some concluding remarks.

The work is written in SI units, and we shall use the known constants:
vacuum permittivity $\epsilon_0=8.85\times 10^{-12}$~F/m, vacuum
permeability $\mu_0=4\pi\times 10^{-7}$~H/m, free space
impedance $\eta_0=\sqrt{\mu_0/\epsilon_0}=376.73\,\Omega$ and
speed of light in vacuum
$c=1/\sqrt{\mu_0\epsilon_0}\approx 3\times 10^8$~m/sec.

\section{Configuration and formulation}

A total amount of charge $q$ is rotating in a circle of radius $d$ at
constant speed $v$ so that the angular velocity is $\omega=v/d$. The
charge $q$ is ``split'' into $N$ charges of value $q/N$, uniformly
distributed around the circle, so that the charge number $k$ is at
the angle $\omega t+2\pi k/N$, where $k=0,1, .. N-1$.  

\begin{figure}[tbp]
\includegraphics[width=18cm]{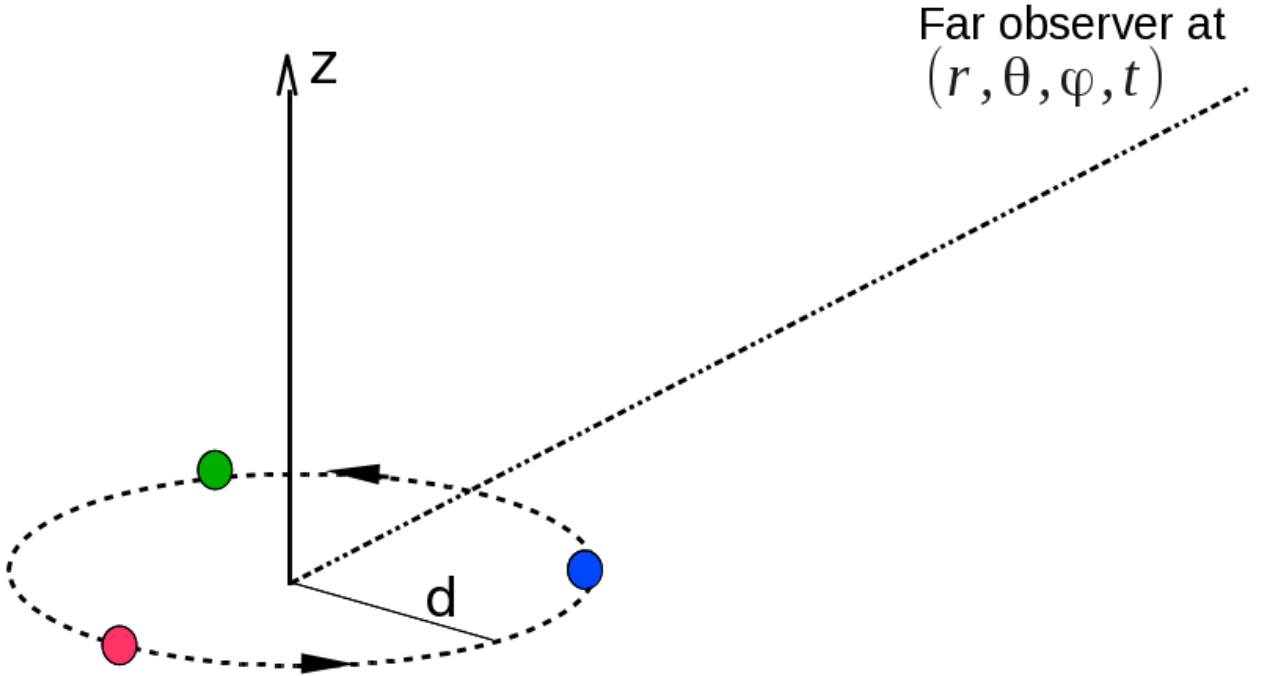}
\caption{(color online) $N$ charges (here $N=3$) of magnitude $q/N$ each, in circular
motion on the $xy$ plane at radius $d$ around the $z$ axis. The far observer's
spherical coordinates are also shown in the figure.}
\label{config}
\end{figure}

The configuration is shown in Figure~\protect\ref{config}, for $N=3$.

The location of the charge $k$ as function of time is given by:

\begin{equation}
\mathbf{r}_k'(t)=d[\mathbf{\widehat{x}} \cos(\omega t+2\pi k/N)+\mathbf{\widehat{y}} \sin(\omega t+2\pi k/N)]
\label{rk}
\end{equation}

The fields propagate with the speed of light $c$. Hence the
fields at the observer location $\mathbf{r}$ at time $t$ are influenced by
the motion of each charge, at an {\it earlier} (retarded) time. Specifically, the fields
are influenced by the motion of the charge $k$ at time $t_k'$ so that

\begin{equation}
R_k \equiv |\mathbf{r}-\mathbf{r}_k'(t_k')|=c(t-t_k')
\label{Rk}
\end{equation}

At large distance from the charges, one may approximate:

\begin{equation}
R_k \approx r-d \sin\theta \cos\phi_k
\label{Rk_far}
\end{equation}

where

\begin{equation}
\phi_k\equiv\omega t_k'+2\pi k/N-\varphi ,
\label{phik}
\end{equation}

hence the retarded time $t_k'$ may be calculated from the following
implicit equation:

\begin{equation}
t_k' = t-r/c + (d/c) \sin\theta \cos\phi_k ,
\label{tk}
\end{equation}

which may be solved numerically by setting a ``1st guess'' $t_k' =
t-r/c$ in the right side of eq.~(\protect\ref{tk}) and recalculate
$t_k'$ until convergence is obtained.

\begin{figure}[tbp]
\includegraphics[width=18cm]{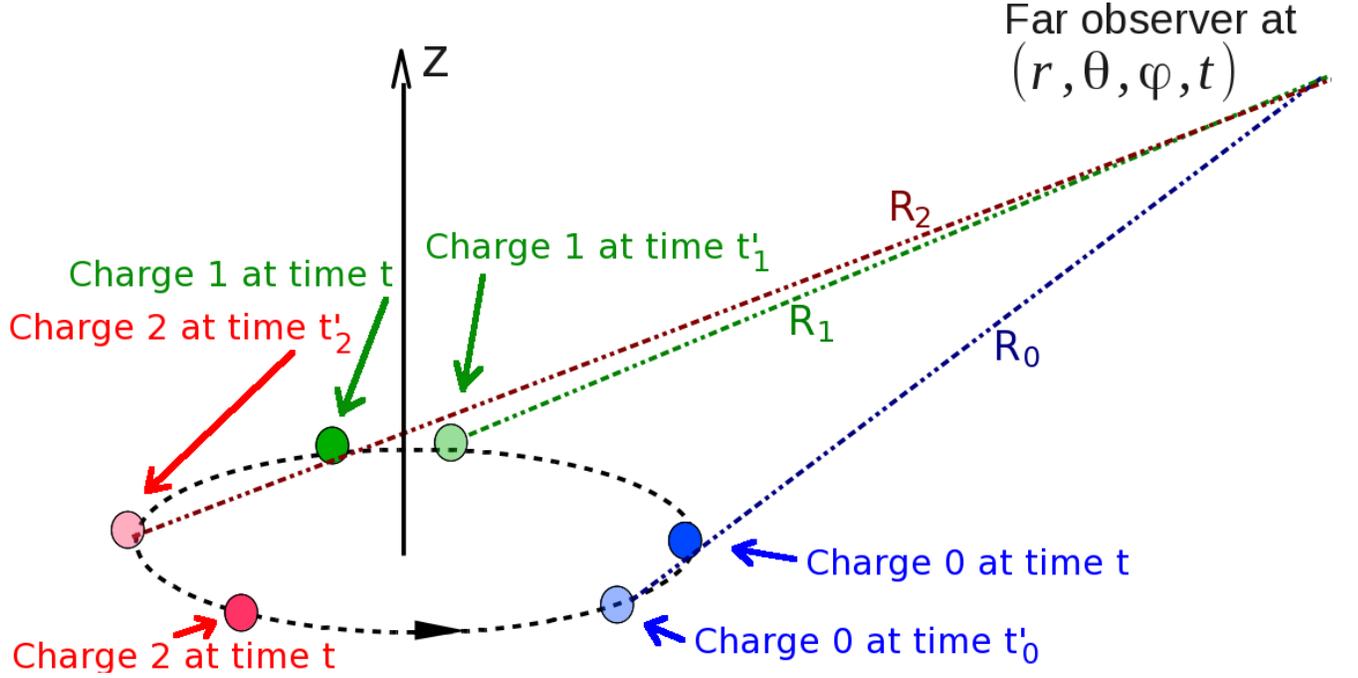}
\caption{(color online) The dark colored spheres represent the charges
at the current position at time $t$ and the light colored spheres
represent the charges at the retarded positions at times $t_k'$. The
retarded positions are connected with dashed lines to the observer
location, and those distances are called $R_0$, $R_1$ and $R_2$. This
emphasizes that the field at observer is determined by the
velocities and accelerations of the charges at the retarded times.}
\label{retarded}
\end{figure}

Figure~\protect\ref{retarded} emphasizes the meaning of retarded
positions of the charges.

Our purpose is to calculate the power radiated by those rotating charges. Of course,
for $N=1$ the result is known by Larmor formula (and will be confirmed later on).

The general form of the Larmor formula
\protect\cite{dirac,rhorlich,ianc_hor,gupta,ianc_hor1,ianc_hor2,bernorz,rhorlich1,baylis}
is:

\begin{equation}
P_{\rm{Larmor}}=\frac{q^2 \gamma^6}{6\pi\epsilon_0 c}(\dot\beta^2-
 (\boldsymbol{\beta}\times\boldsymbol{\dot{\beta}})^2),
\label{larmor}
\end{equation}

where

\begin{equation}
\boldsymbol{\beta}\equiv\mathbf{v}/c.
\label{beta}
\end{equation}

and

\begin{equation}
\gamma\equiv 1/\sqrt{1-\beta^2}
\label{gamma}
\end{equation}

If one defines the angle between the velocity and the acceleration as $\alpha$,
one may express
$(\boldsymbol{\beta}\times\boldsymbol{\dot{\beta}})^2=(\beta\dot{\beta}\sin\alpha)^2$
an rewrite

\begin{equation}
P_{\rm{Larmor}}=\frac{q^2 \gamma^6}{6\pi\epsilon_0 c}\dot\beta^2(1-\beta^2\sin^2\alpha).
\label{larmor1}
\end{equation}

In our case of circular motion, the velocity is perpendicular to the acceleration, so
that $\sin^2\alpha=1$. Therefore by replacing $1-\beta^2=1/\gamma^2$, the Larmor formula
simplifies for our case to

\begin{equation}
\left. P\right|_{\,N=1}=\frac{q^2 \gamma^4\dot\beta^2}{6\pi\epsilon_0 c}=
\frac{q^2 \gamma^4 a^2}{6\pi\epsilon_0 c^3},
\label{larmor2}
\end{equation}

where $a$ is the acceleration. We will calculate how this power decreases when the
number of charges $N$ increases.

\section{Fields and power calculation}

To calculate the power radiated from the collection of charges in
Figure~\protect\ref{config}, one needs only the far fields,
i.e. those who behave like $1/R$. The far electric and magnetic fields
due to the moving charge
$k$ are given by \protect\cite{ianc_hor,rhorlich}:

\begin{equation}
\mathbf{E}_k=(q/N) \mu_0 \frac{\mathbf{\widehat{R}}_k\times[(\mathbf{\widehat{R}}_k-\boldsymbol{\beta}_k)\times\mathbf{a}_k]} {4\pi(1-\boldsymbol{\beta}_k\cdot\mathbf{\widehat{R}}_k)^3 R_k}
\label{Ek}
\end{equation}

and

\begin{equation}
\mathbf{H}_k=\mathbf{\widehat{R}}_k \times \mathbf{E}_k/\eta_0
\label{Hk}
\end{equation}

where $\mathbf{\widehat{R}}_k$ is the unit vector pointing from the
position of the charge to observer and $R_k$ is the distance between the
charge and the observer, as defined in eq.~(\protect\ref{Rk}) - see
Figure~\protect\ref{retarded}. $\boldsymbol{\beta}_k=\mathbf{v}_k/c=\mathbf{\dot{r}}_k'/c$ is the
velocity relative to $c$ and $\mathbf{a}_k=\mathbf{\dot{v}}_k$ is the
acceleration of the charge. All the dynamical variables are evaluated at
the retarded time (defined in eq.~(\protect\ref{tk})).

Defining $\mathbf{\widehat{r}}$ as the unit vector pointing from the coordinates
origin to the observer, we may calculate in the far field the difference:

\begin{equation}
\mathbf{\widehat{R}}_k-\mathbf{\widehat{r}}=\frac{\mathbf{r}-\mathbf{r}_k'}{|\mathbf{r}-\mathbf{r}_k'|}-\frac{\mathbf{r}}{r}
\approx -\frac{\mathbf{r}_k'}{r}
\label{Rkminus_r}
\end{equation}

Hence one may use $\mathbf{\widehat{r}}$ instead of $\mathbf{\widehat{R}}_k$ in eqs.~(\protect\ref{Ek})
and (\protect\ref{Hk}) with an error of order $1/R^2$, which does not affect the
calculations of the radiated power. Also, in the denominator of eq.~(\protect\ref{Ek})
we may set $R_k=r$, as always done for far field. So we express the electric and
magnetic fields as the sum of the contribution from all the charges:

\begin{equation}
\mathbf{E}=\frac{\mu_0 q}{4\pi r N}  \sum_{k=0}^{N-1} \frac{\mathbf{\widehat{r}}\times[(\mathbf{\widehat{r}}-\boldsymbol{\beta}_k)\times\mathbf{a}_k]} {(1-\boldsymbol{\beta}_k\cdot\mathbf{\widehat{r}})^3}
\label{E}
\end{equation}

and

\begin{equation}
\mathbf{H}=\mathbf{\widehat{r}} \times \mathbf{E}/\eta_0
\label{H}
\end{equation}

Now using eq.~(\protect\ref{rk}), we evaluate
$\mathbf{\widehat{r}}\times[(\mathbf{\widehat{r}}-\boldsymbol{\beta}_k)\times\mathbf{a}_k]$
and express it in spherical coordinates:

\begin{equation}
\mathbf{\widehat{r}}\times[(\mathbf{\widehat{r}}-\boldsymbol{\beta}_k)\times\mathbf{a}_k=
a[\boldsymbol{\widehat{\theta}}\cos\theta\cos\phi_k+\boldsymbol{\widehat{\varphi}}(\beta\sin\theta+\sin\phi_k)]
\label{r_cross_r_cross_a}
\end{equation}

and $\boldsymbol{\beta}_k\cdot\mathbf{\widehat{r}}$ evaluates to

\begin{equation}
\boldsymbol{\beta}_k\cdot\mathbf{\widehat{r}}=-\beta\sin\theta\sin\phi_k.
\label{beta_dot_r}
\end{equation}

Setting those results into eq.~(\protect\ref{E}), we obtain

\begin{equation}
\mathbf{E}=\frac{\mu_0 q a}{4\pi r N} [\boldsymbol{\widehat{\theta}}\cos\theta\,\text{Fc}+
\boldsymbol{\widehat{\varphi}}\,\text{Fs}]
\label{E1}
\end{equation}

and

\begin{equation}
\mathbf{H}=\frac{q a}{4\pi r c N} [-\boldsymbol{\widehat{\theta}}\hspace{1mm}\text{Fs}+\boldsymbol{\widehat{\varphi}}\cos\theta\hspace{1mm}\text{Fc}]
\label{H1}
\end{equation}

where the functions Fc and Fs are defined as:

\begin{equation}
\text{Fc}(t,\varphi,\theta,\beta,N)\equiv\sum_{k=0}^{N-1}\text{fc}(\phi_k)
\label{Fc}
\end{equation}

and

\begin{equation}
\text{Fs}(t,\varphi,\theta,\beta,N)\equiv\sum_{k=0}^{N-1}\text{fs}(\phi_k)
\label{Fs}
\end{equation}

and the functions fc and fs are defined as

\begin{equation}
\text{fc}(\phi_k)\equiv\frac{\cos\phi_k}{(1+p\sin\phi_k)^3},
\label{fc}
\end{equation}

and

\begin{equation}
\text{fs}(\phi_k)\equiv\frac{p+\sin\phi_k}{(1+p\sin\phi_k)^3},
\label{fs}
\end{equation}

and $p$ is defined as

\begin{equation}
p\equiv\beta\sin\theta
\label{pp}
\end{equation}

and is a parameter which controls the behavior of $\phi_k$, as will be
soon shown.

The power per unit of normal area (or Poynting vector) is given by,
$\mathbf{S}=\mathbf{E} \times \mathbf{H} = \mathbf{\widehat{r}} E^2/\eta_0$ which results in

\begin{equation}
\mathbf{S}=\mathbf{\widehat{r}} \left[\frac{q a}{4\pi c N r}\right]^2 \eta_0\left|\boldsymbol{\widehat{\varphi}}\hspace{1mm}\text{Fs}+\boldsymbol{\widehat{\theta}}\cos\theta\hspace{1mm}\text{Fc}\right|^2
\label{S}
\end{equation}

The total power is calculated via

\begin{equation}
P= r^2 \int_0^{2\pi}d\varphi \int_0^{\pi}d\theta\sin\theta\, \mathbf{S} \cdot \mathbf{\widehat{r}}
\label{P}
\end{equation}

which results in

\begin{equation}
P=\left. P\right|_{\,N=1}G(t,\beta,N)=\frac{q^2 a^2 \gamma^4}{6\pi\epsilon_0 c^3} G(t,\beta,N).
\label{P1}
\end{equation}

Here we factored out the Larmor formula for the radiation of a single charge - see
eq.~(\protect\ref{larmor2}), so that $G(t,\beta,N)$ is dimensionless and represents
the decay of the power. The function $G(t,\beta,N)$ is given by

\begin{equation}
G(t,\beta,N)\equiv\frac{3}{8\pi\gamma^4 N^2} \int_0^{2\pi}d\varphi \int_0^{\pi}d\theta\sin\theta\hspace{1mm} F(t,\varphi,\theta,\beta,N)
\label{G}
\end{equation}

hence $G=1$ for $N=1$, for any $t$ or $\beta$ and the function $F$ is

\begin{equation}
F(t,\varphi,\theta,\beta,N)\equiv\left|\boldsymbol{\widehat{\varphi}}\hspace{1mm}\text{Fs}+\boldsymbol{\widehat{\theta}}\cos\theta\hspace{1mm}\text{Fc}\right|^2= \text{Fs}^2 + \cos^2\theta\hspace{1mm} \text{Fc}^2
\label{F}
\end{equation}

Now we eliminate $t_k'$ from eq.~(\protect\ref{tk}) and rewrite the implicit
eqs.~(\protect\ref{tk}) and (\protect\ref{phik}) in terms of $\phi_k$

\begin{equation}
\phi_k=\omega(t-r/c)-\varphi+2\pi k/N+p\cos\phi_k
\label{phik1}
\end{equation}

This allows us
to change variable $\varphi'=\varphi-\omega(t-r/c)$ in (\protect\ref{G})
obtaining:

\begin{equation}
G(t,\beta,N)=\frac{3}{8\pi\gamma^4  N^2} \int_{-\omega(t-r/c)}^{2\pi-\omega(t-r/c)}d\varphi' \int_0^{\pi}d\theta\sin\theta\hspace{1mm} (\text{Fs}^2+\cos^2\theta\hspace{1mm}\text{Fc}^2),
\label{G1}
\end{equation}

so that eq.~(\protect\ref{phik1}) is rewritten as

\begin{equation}
\phi_k=-\varphi'+2\pi k/N+p\hspace{1mm}\cos\phi_k ,
\label{phik2}
\end{equation}

We see that if $\phi_k$ and $\varphi'$ satisfy eq.~(\protect\ref{phik2}), also
$\phi_k-2\pi$ and $\varphi'+2\pi$ satisfy it, hence $\cos\phi_k$ and
$\sin\phi_k$ are periodic functions of $\varphi'$, with a periodicity of $2\pi$. 
Therefore, the $d\varphi'$ integral in eq.~(\protect\ref{G1}) may be evaluated
over any period of $2\pi$, showing that $G$ (and therefore also the radiated power
$P$) does {\it not depend on time}, so that we may simplify eq.~(\protect\ref{G1}) to

\begin{equation}
G(\beta,N)=\frac{3}{8\pi\gamma^4  N^2}  \int_0^{2\pi}d\varphi' \int_0^{\pi}d\theta\sin\theta\hspace{1mm} (\text{Fs}^2+\cos^2\theta\hspace{1mm}\text{Fc}^2).
\label{G2}
\end{equation}

We redefined $G$ to be time independent, so that fc, fs, Fc and Fs in eqs.~(\protect\ref{fc}),
(\protect\ref{fs}), (\protect\ref{Fc}) and (\protect\ref{Fs}) become functions of
$\varphi'$ instead of $t$ and $\varphi$.

For understanding the behavior of Fs and Fc, we plot fc, fs in
eqs.~(\protect\ref{fc}) and (\protect\ref{fs}), and theirs sums
(eqs.~(\protect\ref{Fc}) and (\protect\ref{Fs}))
for different parameters.

\begin{figure}[htbp]
\includegraphics[width=8cm]{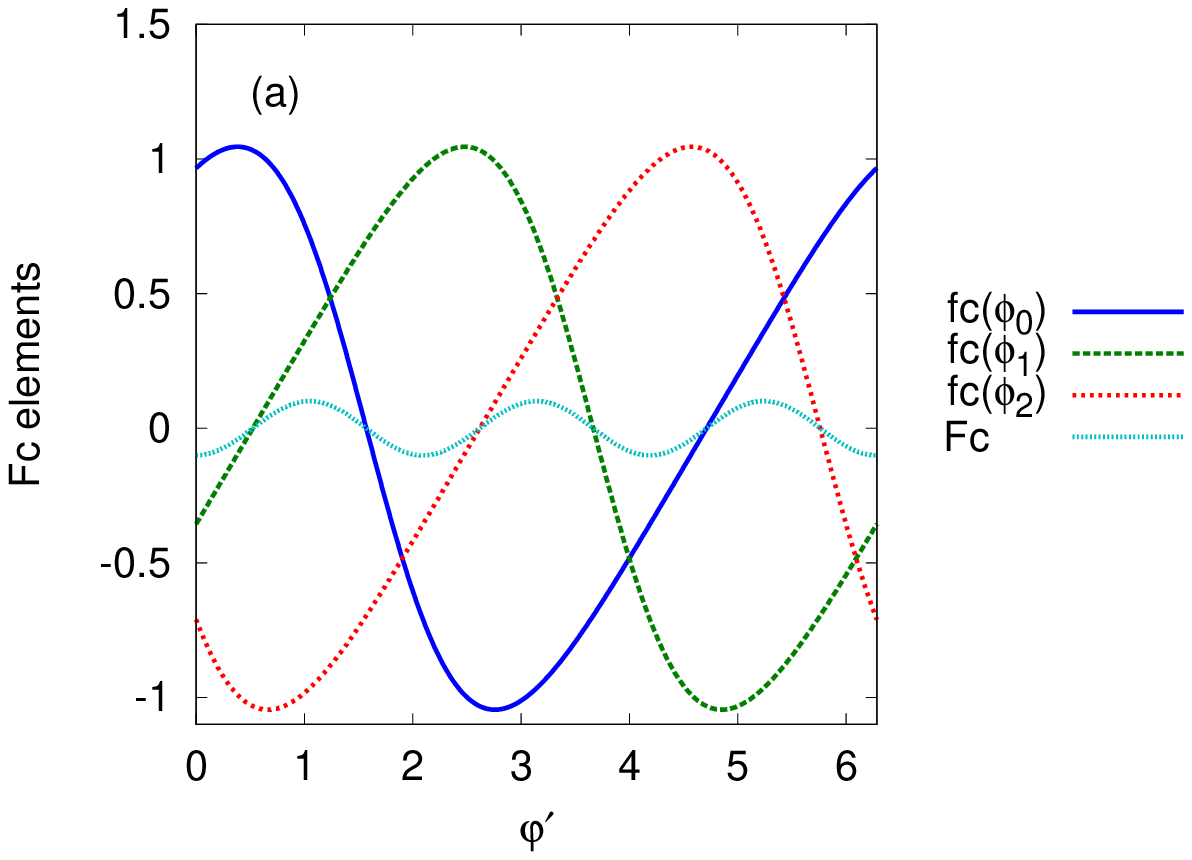}
\includegraphics[width=8cm]{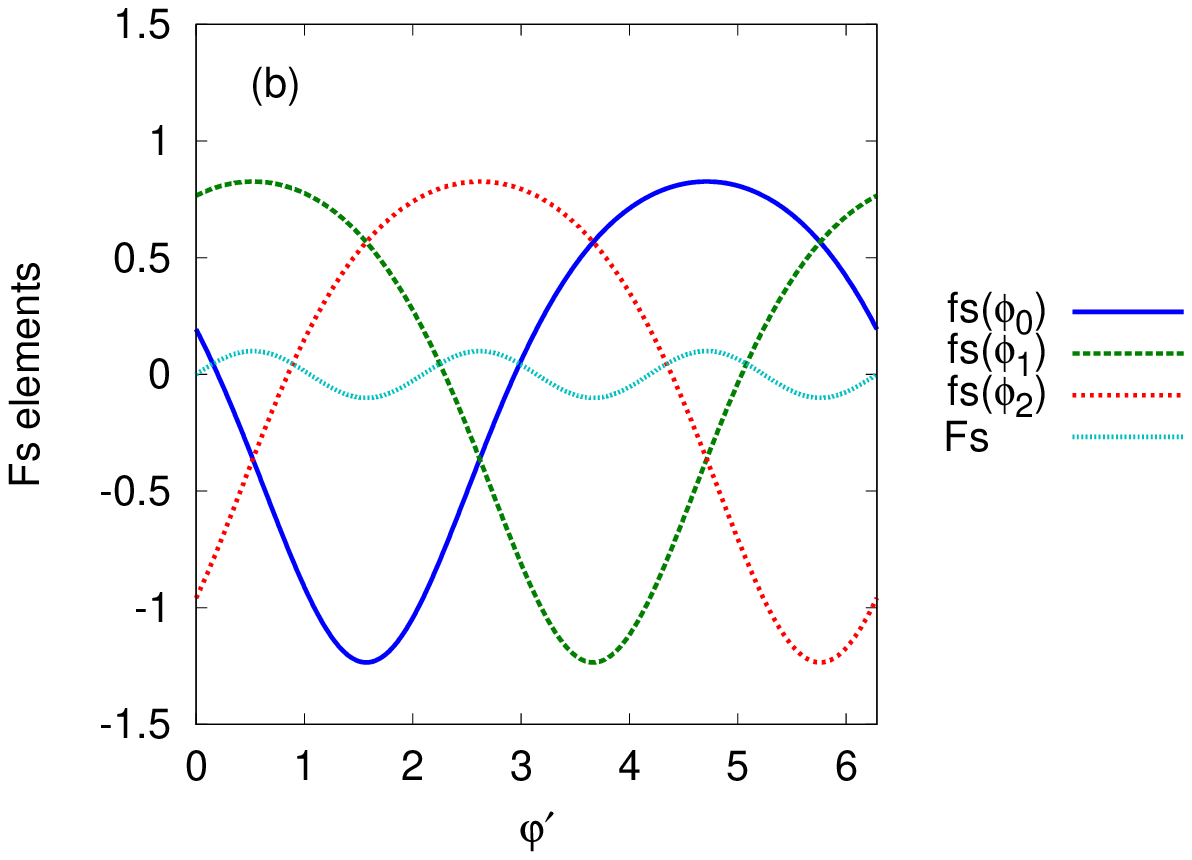}
\caption{(color online) The fc in eq.~(\protect\ref{fc}) and their sum
is shown in panel (a) and the fs in eq.~(\protect\ref{fs}) and their sum
is shown in panel (b) for $N=3$ and $p=0.1$.}
\label{sin_cos_01_3}
\end{figure}

\begin{figure}[htbp]
\includegraphics[width=8cm]{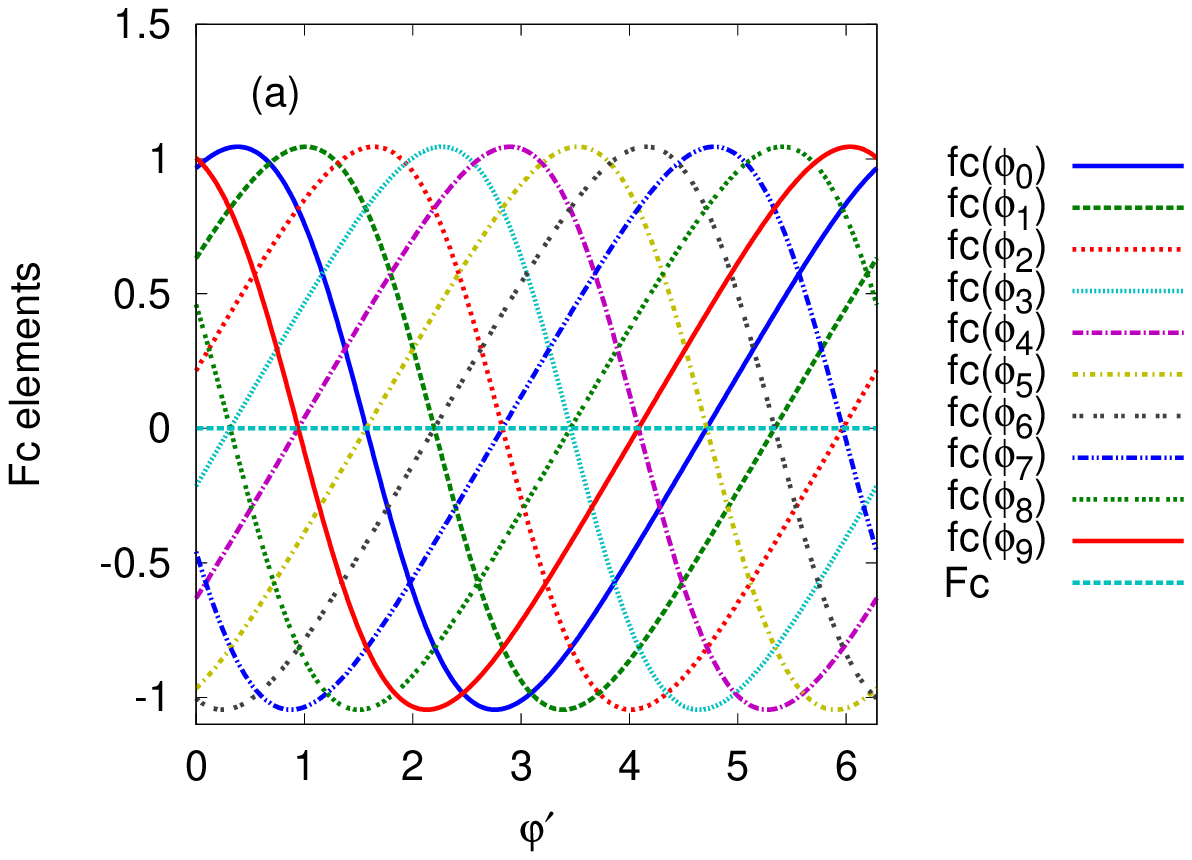}
\includegraphics[width=8cm]{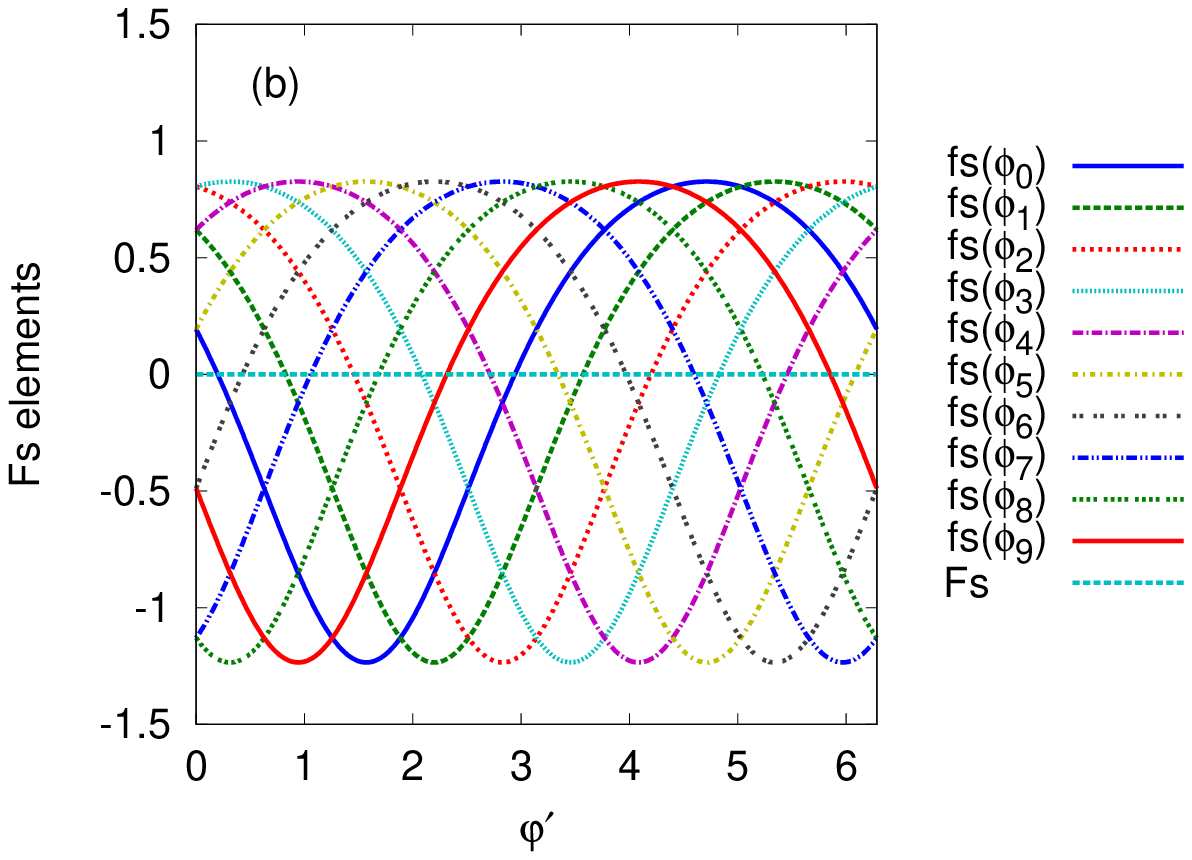}
\caption{(color online) The fc in eq.~(\protect\ref{fc}) and their sum
is shown in panel (a) and the fs in eq.~(\protect\ref{fs}) and their sum
is shown in panel (b) for $N=10$ and $p=0.1$.}
\label{sin_cos_01_10}
\end{figure}

Clearly for very small $p$ in eq.~(\protect\ref{phik2}),
$\phi_k\approx -\varphi'+2\pi k/N$, hence the cosine or sine of $\phi_k$
equal approximately to the cosine or sine of $-\varphi'+2\pi k/N$, so
that both have harmonic shapes as function of $\varphi'$.
In such case $\text{fc}\approx\cos(-\varphi'+2\pi k/N)$ and
$\text{fs}\approx\sin(-\varphi'+2\pi k/N)$, hence they sum
to a small value as observed in Figures~\protect\ref{sin_cos_01_3} and
\protect\ref{sin_cos_01_10}. For $N=3$ (Figure~\protect\ref{sin_cos_01_3})
the amplitudes of Fc and Fs are around 0.1, and this decreases with $N$, 
as may be seen in Figure~\protect\ref{sin_cos_01_10}, for $N=10$.

\begin{figure}[htbp]
\includegraphics[width=8cm]{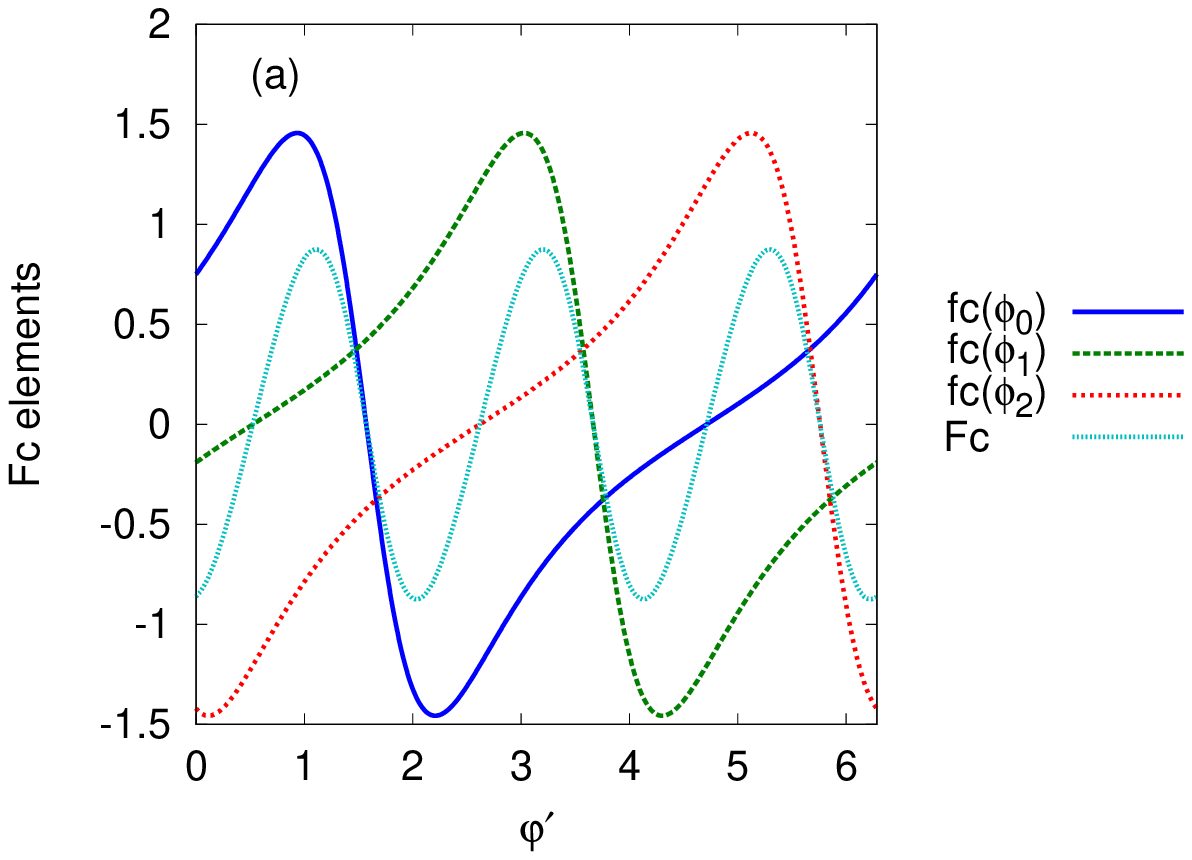}
\includegraphics[width=8cm]{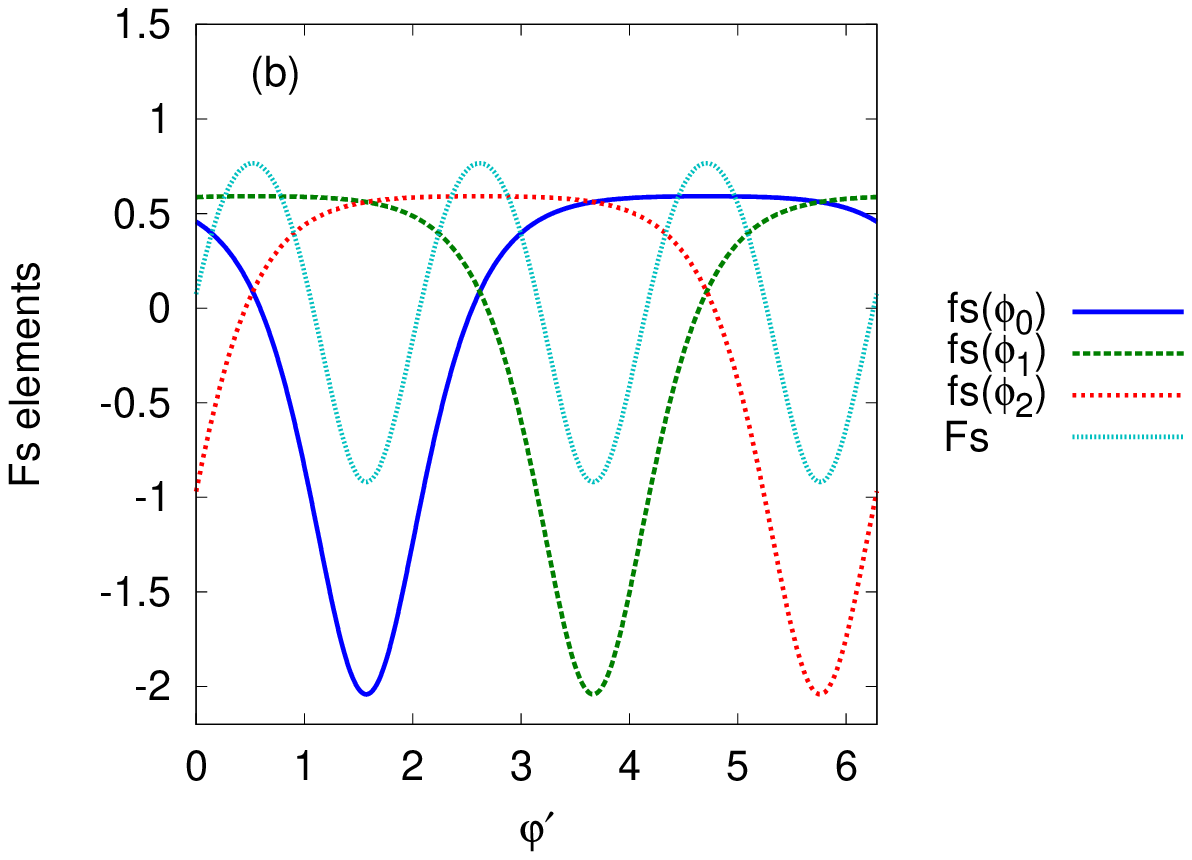}
\caption{(color online) The fc in eq.~(\protect\ref{fc}) and their sum
is shown in panel (a) and the fs in eq.~(\protect\ref{fs}) and their sum
is shown in panel (b) for $N=3$ and $p=0.3$.}
\label{sin_cos_03_3}
\end{figure}

\begin{figure}[htbp]
\includegraphics[width=8cm]{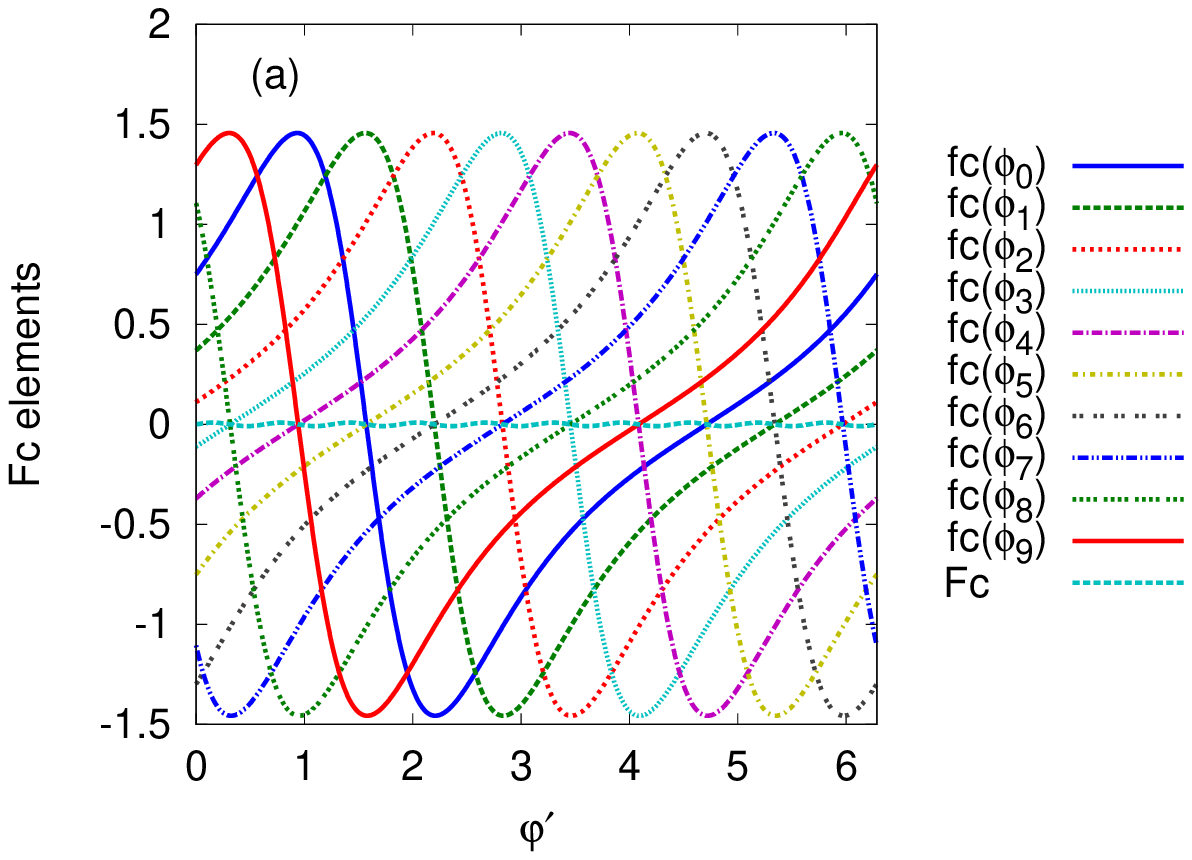}
\includegraphics[width=8cm]{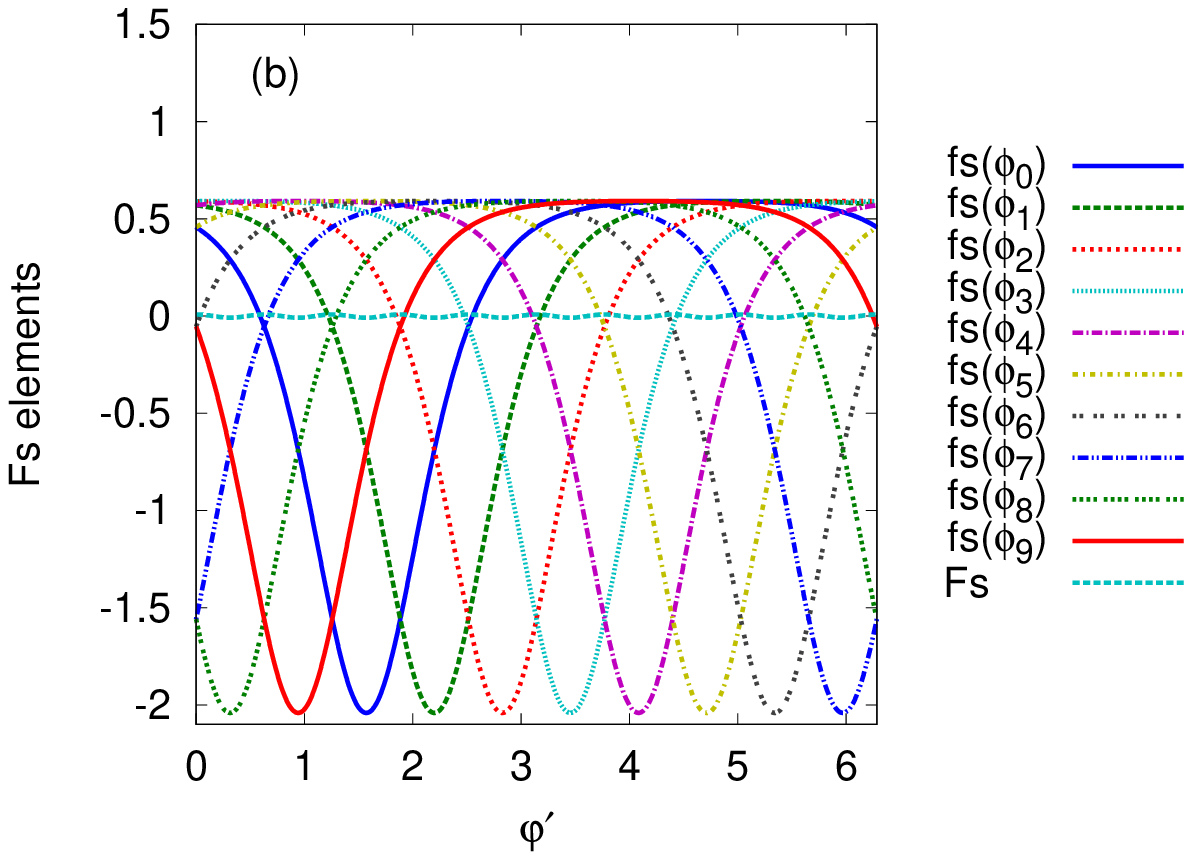}
\caption{(color online) The fc in eq.~(\protect\ref{fc}) and their sum
is shown in panel (a) and the fs in eq.~(\protect\ref{fs}) and their sum
is shown in panel (b) for $N=10$ and $p=0.3$.}
\label{sin_cos_03_10}
\end{figure}

\begin{figure}[htbp]
\includegraphics[width=8cm]{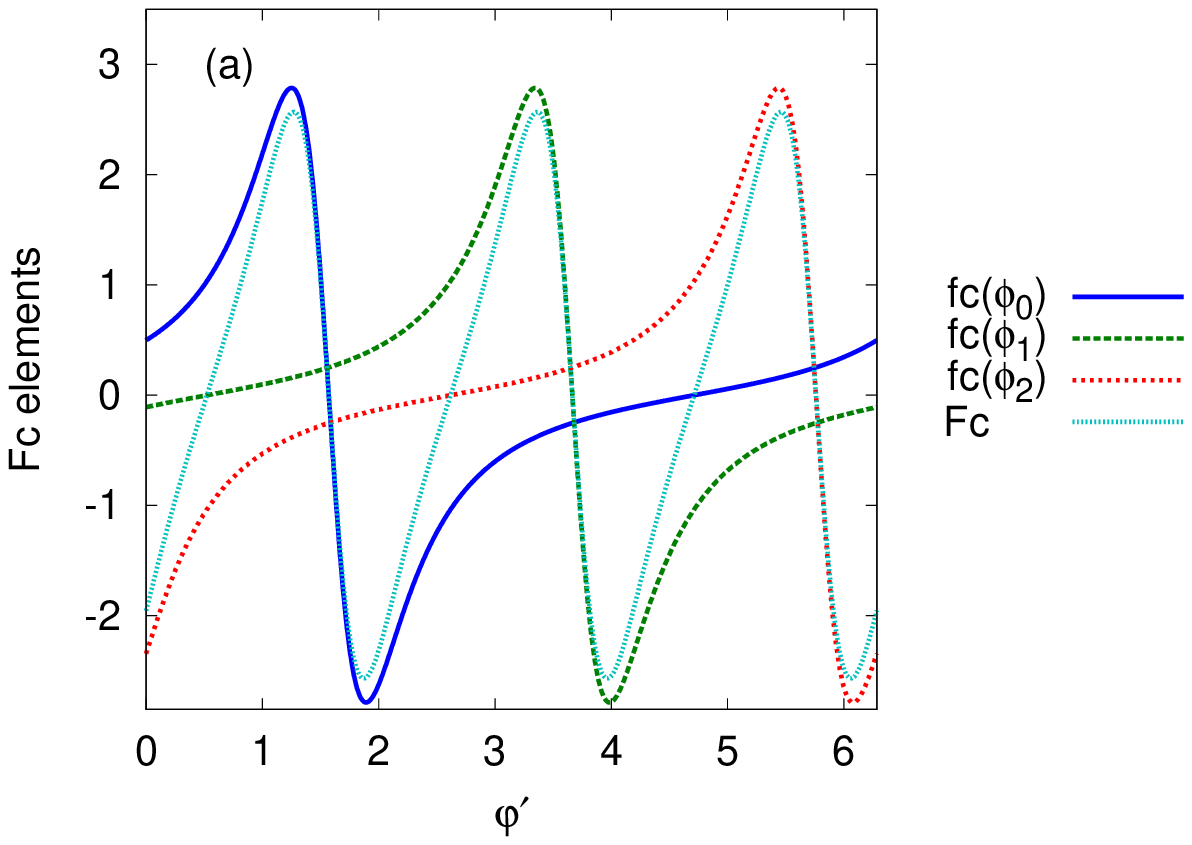}
\includegraphics[width=8cm]{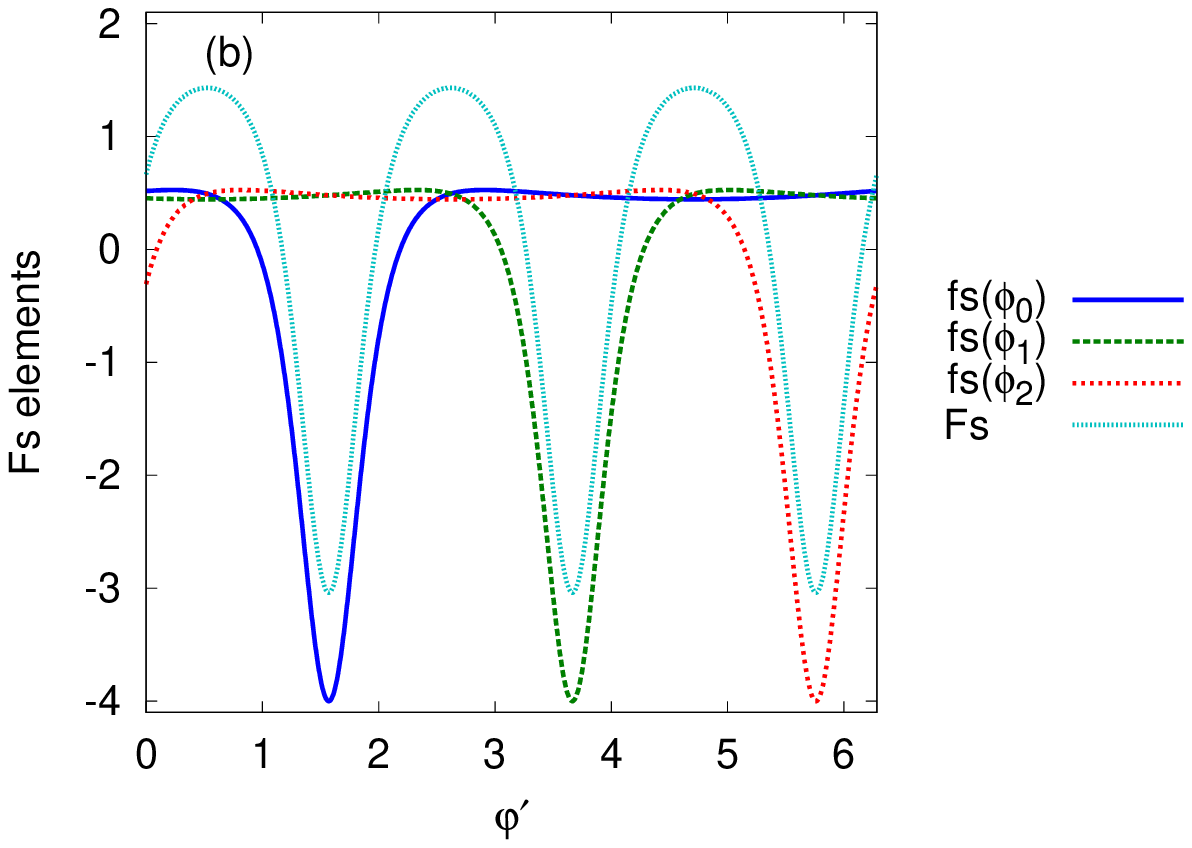}
\caption{(color online) The fc in eq.~(\protect\ref{fc}) and their sum
is shown in panel (a) and the fs in eq.~(\protect\ref{fs}) and their sum
is shown in panel (b) for $N=3$ and $p=0.5$.}
\label{sin_cos_05_3}
\end{figure}

\begin{figure}[htbp]
\includegraphics[width=8cm]{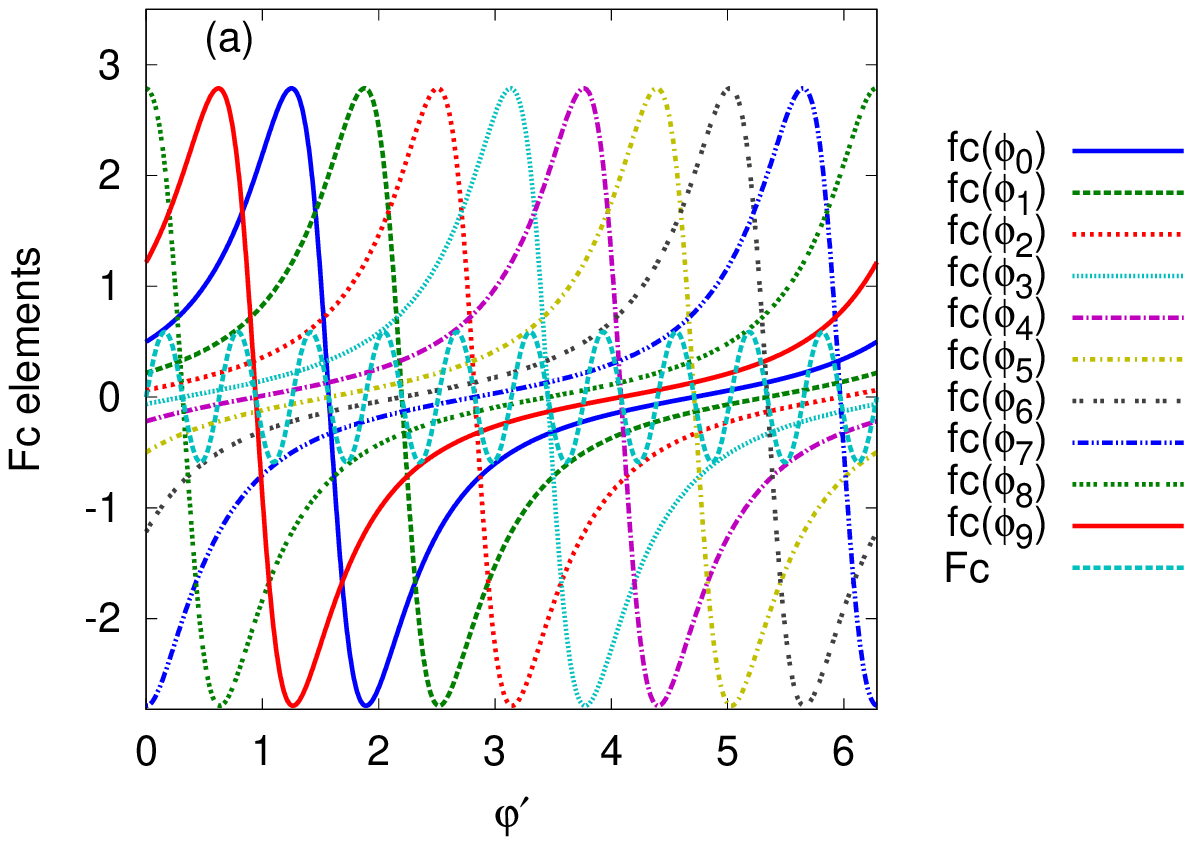}
\includegraphics[width=8cm]{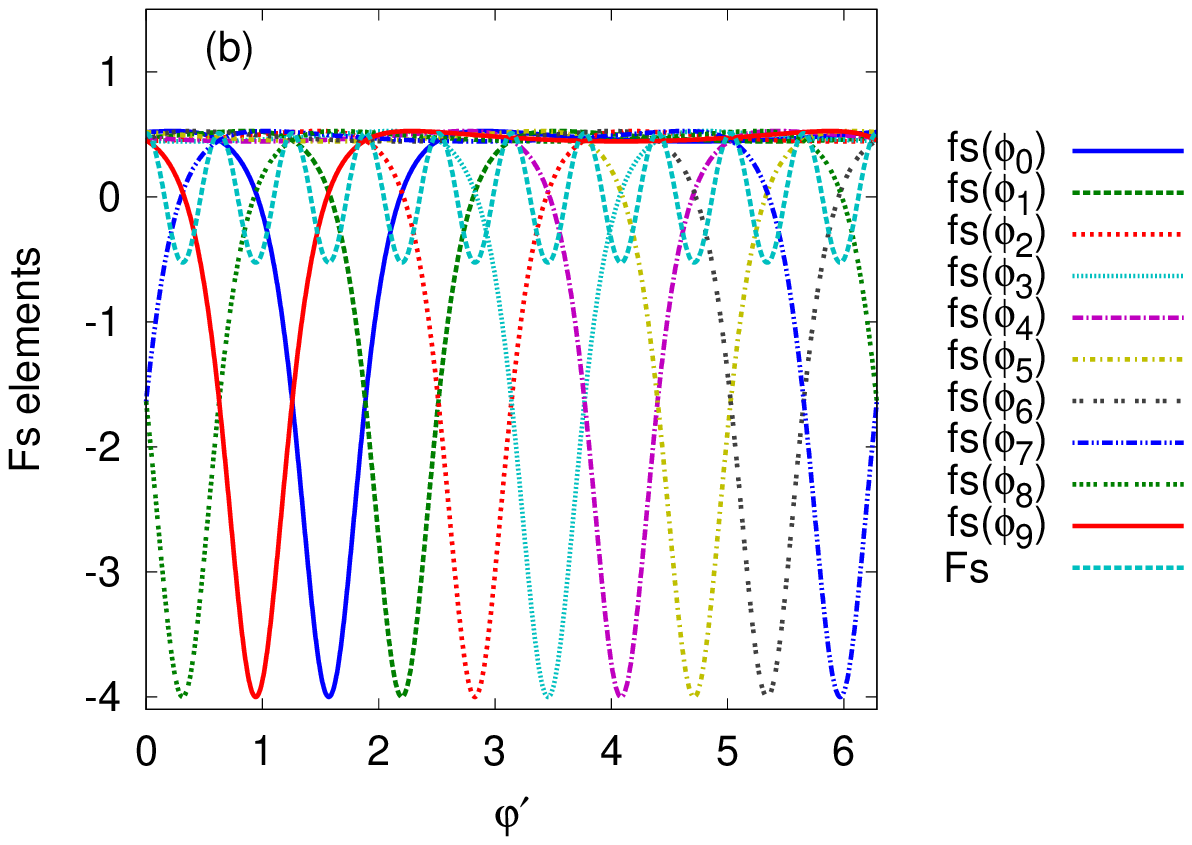}
\caption{(color online) The fc in eq.~(\protect\ref{fc}) and their sum
is shown in panel (a) and the fs in eq.~(\protect\ref{fs}) and their sum
is shown in panel (b) for $N=10$ and $p=0.5$.}
\label{sin_cos_05_10}
\end{figure}

As $p$ increases, fc and fs in eqs.~(\protect\ref{fc})
and (\protect\ref{fs}) get more distorted, hence they sum to bigger
amplitudes, as observed in Figures~\protect\ref{sin_cos_03_3},
\protect\ref{sin_cos_03_10}, \protect\ref{sin_cos_05_3} and
\protect\ref{sin_cos_05_10}.

In Figure~\protect\ref{sin_cos_03_3}, showing the behavior for $p=0.3$
and $N=3$, Fc and Fs have amplitudes of 0.87 and 0.76, respectively, and
those decrease for $N=10$ (Figure~\protect\ref{sin_cos_03_10}) to 
0.0086 and 0.0082. Also we see that for $p=0.3$, although fc and fs
are distorted, the sums Fc and Fs are almost undistorted, unlike for
the $p=0.5$ and $N=3$ case in Figure~\protect\ref{sin_cos_05_3}.

The case of $p=0.5$ is shown in Figures~\protect\ref{sin_cos_05_3} and
\protect\ref{sin_cos_05_10}. In this case, not only fc and fs are
distorted, but also the sums Fc and Fs, however the distortion of the
sum decreases when the number of charges $N$ increases, as may be seen
for the case $N=10$ in Figure~\protect\ref{sin_cos_05_10}. In the last
case the amplitudes of Fc and Fs are 0.59 and 0.5, much bigger than
in the parallel case with $p=0.3$ .

To summarize, the functions Fc and Fs increase with $p$ and decrease with
$N$, tending to undistorted harmonic functions, for big values of $N$.

It is interesting to remark that the average of the functions fc and fs
is always 0, although this is not always visible for the fs functions.
This may be proved by calculating

\begin{equation}
\langle\text{fc,s}\rangle = \frac{1}{2\pi}\int_{0}^{2\pi}d\varphi'\text{fc,s}(\phi_k)=
\frac{1}{2\pi}\int_{0}^{2\pi}d\varphi'\frac{\text{hc,s}(\phi_k)}{(1+p\sin\phi_k)^3}
\label{Avfc}
\end{equation}

where for brevity we called $\text{fc,s}$ the functions fc or fs, and we called their average
$\langle\text{fc,s}\rangle$. We also use the abbreviation $\text{hc,s}$ for the functions
hc and hs defined as

\begin{equation}
\text{hc}(\phi_k)\equiv\cos\phi_k
\label{hc}
\end{equation}

\begin{equation}
\text{hs}(\phi_k)\equiv p+\sin\phi_k
\label{hs}
\end{equation}

for the fc and fs average calculation, respectively. We change variable from $\varphi'$
to $\phi_k$, and we find from eq.~(\protect\ref{phik2}) that

\begin{equation}
d\varphi'/d\phi_k=-1-p\hspace{1mm}\sin\phi_k ,
\label{dphi_over_dphik}
\end{equation}

getting:

\begin{equation}
\langle\text{fc,s}\rangle=-\frac{1}{2\pi}\int_{\phi_k(0)}^{\phi_k(0)-2\pi}d\phi_k
\frac{\text{hc,s}(\phi_k)}{(1+p\sin\phi_k)^2} ,
\label{Avfc1}
\end{equation}

where $\phi_k(0)$ is the value of $\phi_k$ at $\varphi'=0$. Because the
integrand has a periodicity of $2\pi$, one may integrate over any period
of $2\pi$, obtaining:

\begin{equation}
\langle\text{fc,s}\rangle=\frac{1}{2\pi}\int_{0}^{2\pi}d\phi_k \frac{\text{hc,s}(\phi_k)}{(1+p\sin\phi_k)^2} .
\label{Avfc2}
\end{equation}

\begin{figure}[htbp]
\includegraphics[width=15cm]{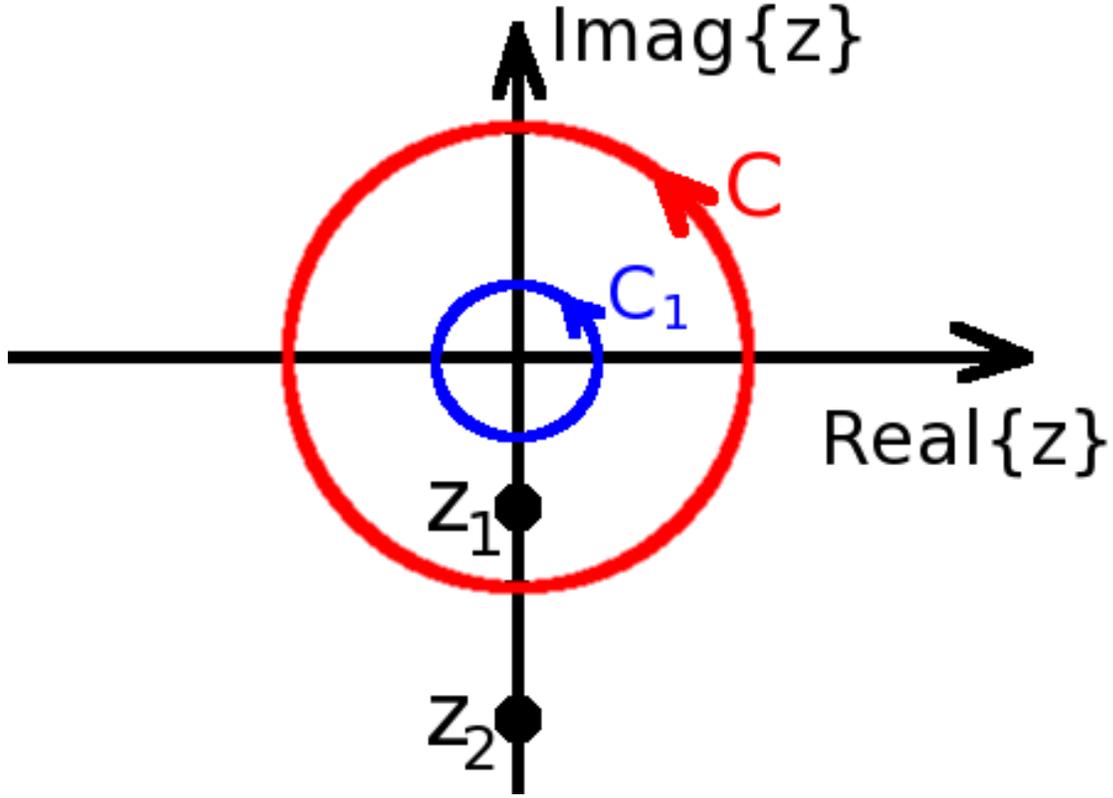}
\caption{(color online) The complex $z$ plane on which we show: the poles $z_{1,2}$ in
eq~(\protect\ref{z12}), the integration contour $C$ used in eqs.~(\protect\ref{Avfc3})
and (\protect\ref{Acs_1_c}), and the integration contour $C_1$ used in
eq.~(\protect\ref{Acs_1_c1}). The poles $z_{1,2}$ are negative pure imaginary,
and $|z_1|<1$ and $|z_2|>1$, the integration contour $C$ is on the unit circle
and the integration contour $C_1$ is on a circle of radius smaller than $|z_1|$.}
\label{poles}
\end{figure}

This integral may be solved by the residue method on the complex plane.
After changing variable $z=\exp(i\phi_k)$, one obtains

\begin{equation}
\langle\text{fc,s}\rangle=\frac{1}{2\pi}\frac{2i}{p^2}\oint_{C}dz
\frac{\text{lc,s}(z)}{(z-z_1)^2 (z-z_2)^2}\equiv \frac{1}{2\pi}\frac{2i}{p^2}\oint_{C}dz\, \text{Lc,s}(z) .
\label{Avfc3}
\end{equation}

where $C$ is the counterclockwise unit circle integration contour shown
in Figure~(\protect\ref{poles}) and $\text{lc,s}$ are abbreviations for

\begin{equation}
\text{lc}(z)\equiv z^2+1
\label{lc}
\end{equation}

and

\begin{equation}
\text{ls}(z)\equiv -i(z^2-4bz-1),
\label{ls}
\end{equation}

where $b$ is a pure imaginary number defined by

\begin{equation}
b\equiv -0.5ip,
\label{b}
\end{equation}

and the 2nd order poles are (expressed in terms of $p$ or $b$)

\begin{equation}
z_{1,2}=-i\left(1/p \mp \sqrt{(1/p)^2-1}\right)=\frac{1}{2}\left(-b^{-1}\pm\sqrt{b^{-2}+4}\right).
\label{z12}
\end{equation}

where indices 1 and 2 refer to upper and lower signs respectively -
see Figure~(\protect\ref{poles}).

In the last expression of the poles in terms of $b$, the magnitude under the square
root is real and negative, so the square root is understood to be positive pure imaginary.

Also we named the integrand in eq.~(\protect\ref{Avfc3}) $\text{Lc,s}$, defined as

\begin{equation}
\text{Lc,s}=\frac{\text{lc,s}(z)}{(z-z_1)^2 (z-z_2)^2}
\label{Lz}
\end{equation}

to ease manipulations. We remark that only $z_1$ is inside the integration
contour (see Figure~(\protect\ref{poles})), and its residue is

\begin{equation}
\left.\text{Res}(L,z1)=\lim_{z\rightarrow z_1}\frac{d}{dz}[\text{Lc,s}(z)(z-z_1)^2]=
\frac{d}{dz}\left[\frac{\text{lc,s}(z)}{(z-z_2)^2}\right]\right|_{ z=z_1}=
\frac{\text{lc,s}'(z_1)(z_1-z_2)-2\text{lc,s}'(z_1)}{(z_1-z_2)^3}.
\label{Res_00}
\end{equation}

Using the relations $z_1 z_2 = -1$ and $z_1+z_2 = -2i/p$, one finds that
for both fc and fs cases the nominator of eq.~(\protect\ref{Res_00}) is 0,
and hence

\begin{equation}
\langle\text{fc}\rangle=\langle\text{fs}\rangle=0 .
\label{Avfc4}
\end{equation}

Now we continue with the calculation of $G$ in eq.~(\protect\ref{G2}).
Rewriting eq.~(\protect\ref{phik2}) for the charge $m$ instead of
$k$ results in

\begin{equation}
\phi_m=-\varphi'+2\pi m/N+p\hspace{1mm}\cos\phi_m ,
\label{phim2}
\end{equation}

which may be rewritten as

\begin{equation}
\phi_m=-(\varphi'-2\pi (m-k)/N)+2\pi k/N+p\hspace{1mm}\cos\phi_m ,
\label{phim3}
\end{equation}

showing that if we could explicitly express $\phi_k(\varphi')$ from
eq.~(\protect\ref{phik2}), $\phi_m$ would be the {\it same} function
of $\varphi'$ only shifted:

\begin{equation}
\phi_k(\varphi')=\phi_m(\varphi'-2\pi (m-k)/N)
\label{phik_m}
\end{equation}

as also evident from Figures~\protect\ref{sin_cos_01_3}~-~\protect\ref{sin_cos_05_10}.
Therefore both Fs and Fc functions remain unchanged for a shift of
multiples of $2\pi/N$, say:

\begin{equation}
\text{Fc}(\varphi'+2\pi/N),\theta,\beta,N)=\sum_{k=0}^{N-1}\frac{\cos\phi_k(\varphi'+2\pi/N)}{(1+p\sin\phi_k(\varphi'+2\pi/N))^3}=\sum_{k=0}^{N-1}\frac{\cos\phi_{k+1}(\varphi')}{(1+p\sin\phi_{k+1}(\varphi'))^3} .
\label{Fc_invariance}
\end{equation}

and the sum being on all charges (and $k$ is modulo $N$), we are left with the same
result. We may therefore rewrite in eq.~(\protect\ref{G2})

\begin{equation}
\int_0^{2\pi}d\varphi'=\sum_{n=0}^{N-1}\int_{2\pi n/N}^{2\pi(n+1)/N}d\varphi'
\label{split_integ}
\end{equation}

and after changing variable $\varphi''=\varphi'-2\pi n/N$, we are left with $N$
identical integrals over the period 0 to $2\pi/N$. For simplicity we rename
$\varphi''$ back to $\varphi'$ and rewrite the function $G$ as:

\begin{equation}
G(\beta,N)=\frac{3}{8\pi\gamma^4 N} \int_0^{2\pi/N}d\varphi' \int_0^{\pi}d\theta\sin\theta\hspace{1mm} (\text{Fs}^2+\cos^2\theta\hspace{1mm}\text{Fc}^2),
\label{G3}
\end{equation}

and this will significantly reduce the time of a numerical integration.
Now looking at the $\theta$ dependence of Fs and Fc, we remark from
eq.~(\protect\ref{phik2}) that $\phi_k$ depends on $p=\beta\sin\theta$, hence
it is invariant under replacing $\theta$ by $\pi-\theta$, and so is
$\cos^2\theta$. We may therefore replace the integration from 0 to $\pi$
by twice the integration from 0 to $\pi/2$, getting

\begin{equation}
G(\beta,N)=\frac{3}{4\pi\gamma^4 N} \int_0^{2\pi/N}d\varphi' \int_0^{\pi/2}d\theta\sin\theta\hspace{1mm} (\text{Fs}^2+\cos^2\theta\hspace{1mm}\text{Fc}^2),
\label{G4}
\end{equation}

Now we change to the variable $p$ defined in eq.~(\protect\ref{pp})
and rewrite eq.~(\protect\ref{G4}) obtaining:

\begin{equation}
G(\beta,N)=\frac{3}{4\pi N\gamma^4\beta^2} \int_0^{2\pi/N}d\varphi' \int_0^{\beta} dp\hspace{1mm}
  p \left(\text{Fs}^2/\sqrt{1-(p/\beta)^2}+\text{Fc}^2 \sqrt{1-(p/\beta)^2}\right),
\label{G5}
\end{equation}

For the case of $N=1$ we know $G$ must be 1, but for consistency we shall prove it:

\begin{align}
G(\beta,1)=\frac{3}{4\pi \gamma^4\beta^2} \int_0^{2\pi}d\varphi' \int_0^{\beta} dp\, p & \left(\left[\frac{p+\sin\phi_0}{(1+p\sin\phi_0)^3}\right]^2/\sqrt{1-(p/\beta)^2}\right. +  \notag \\
& \left. \left[\frac{\cos\phi_0}{(1+p\sin\phi_0)^3}\right]^2 \sqrt{1-(p/\beta)^2}\right),
\label{G_1}
\end{align}

where $\text{Fc}^2$ and $\text{Fs}^2$ reduced here to a single term.
By changing integration order, we may perform the $d\varphi'$ integration
by the change of
variable $d\varphi'/d\phi_0$ defined in eq.~(\protect\ref{dphi_over_dphik}),
obtaining

\begin{equation}
G(\beta,1)=\frac{3}{4\pi \gamma^4\beta^2} \int_0^{\beta} dp\, p \int_{\phi_0(0)-2\pi}^{\phi_0(0)}d\phi_0 \left(\frac{(p+\sin\phi_0)^2}{(1+p\sin\phi_0)^5 \sqrt{1-(p/\beta)^2}}+
 \frac{\cos^2\phi_0\sqrt{1-(p/\beta)^2}}{(1+p\sin\phi_0)^5} \right).
\label{G_11}
\end{equation}

All the functions have a $2\pi$ periodicity on $\phi_0$, so one may use
any limits of interval $2\pi$ for $\phi_0$. By changing variable
$z=\exp(i\phi_0)$ and using the residue method on the complex plane
we obtain

\begin{equation}
\int_0^{2\pi}d\phi_0 \frac{\cos^2\phi_0}{(1+p\sin\phi_0)^5}=
\frac{\pi}{4}\frac{4+p^2}{(1-p^2)^{7/2}}
\label{int_cos2}
\end{equation}

and

\begin{equation}
\int_0^{2\pi}d\phi_0 \frac{(p+\sin\phi_0)^2}{(1+p\sin\phi_0)^5}=
\frac{\pi}{4}\frac{4+3 p^2}{(1-p^2)^{5/2}},
\label{int_sin2}
\end{equation}

hence

\begin{equation}
G(\beta,1)=\frac{3}{16\gamma^4\beta^2} \int_0^{\beta} dp\hspace{1mm} p \left[
 \frac{(4+p^2)\sqrt{1-(p/\beta)^2}}{(1-p^2)^{7/2}}+
 \frac{(4+3p^2)}{(1-p^2)^{5/2}\sqrt{1-(p/\beta)^2}},
\right]
\label{G_12}
\end{equation}

which results in

\begin{equation}
G(\beta,1)=\frac{3}{16\gamma^4\beta^2} \left[
 \frac{2\beta^2(2-\beta^2)}{3(1-\beta^2)^2}+
 \frac{2\beta^2(6+\beta^2)}{3(1-\beta^2)^2}
\right]=\frac{3}{16\gamma^4\beta^2} \frac{16\beta^2}{3(1-\beta^2)^2}=1
\label{G_13}
\end{equation}\\

\begin{figure}[htbp]
\includegraphics[width=18cm]{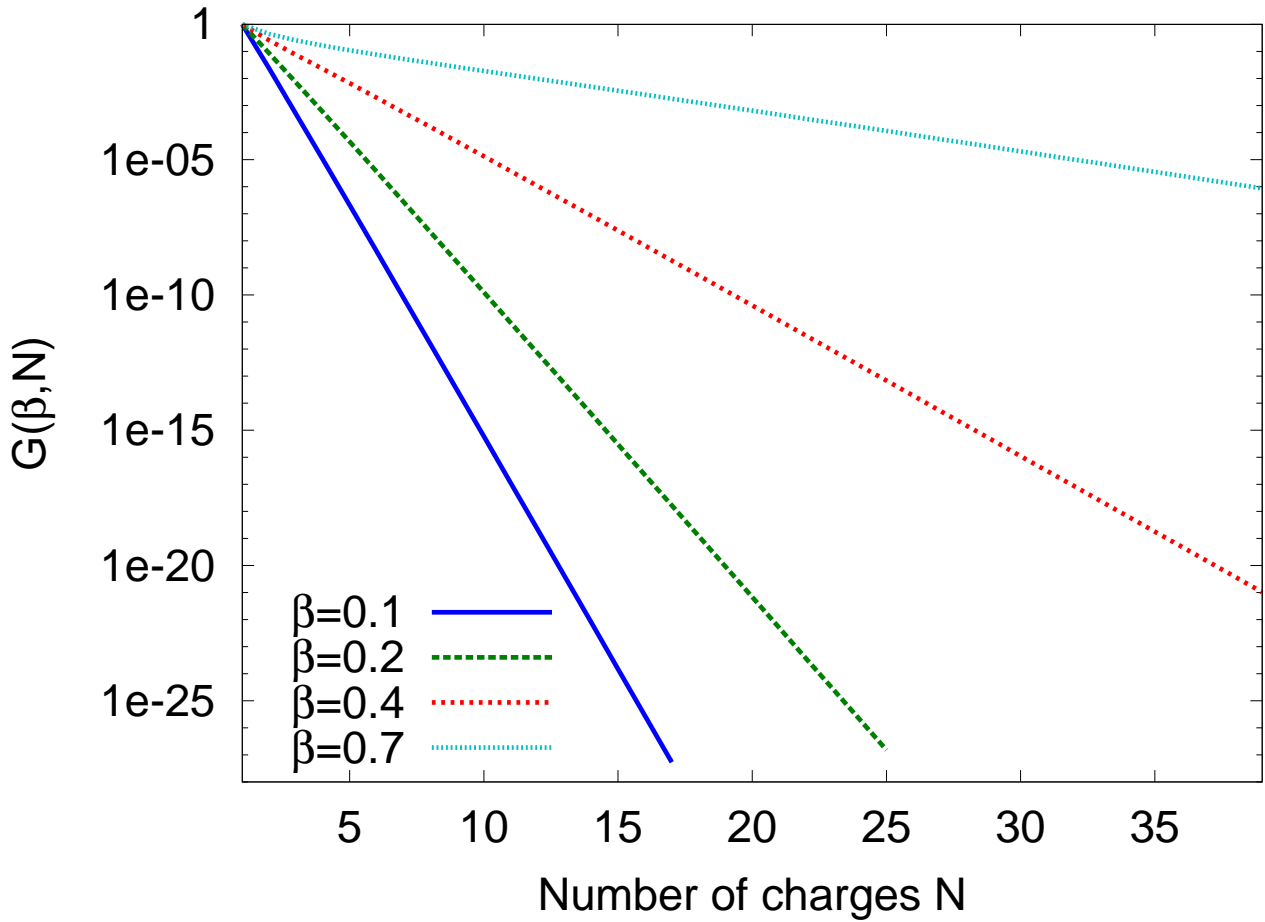}
\caption{Result of $G$ (eq.~(\protect\ref{G5})) as function of the number
of charges $N$, for different values of $\beta$. For big $N$, $G$
goes asymptotically to 0 and as smaller $\beta$ is, $G$ goes faster to 0.}
\label{GFuncN}
\end{figure}

We perform now the calculation in eq.~(\protect\ref{G5}) numerically.
Knowing that $G(\beta,1)=1$, this calculation actually shows
the power radiated by $N$ charges, divided by the power radiated
by a single charge. The results are shown as function of $N$
for different values of $\beta$ in Figure~\protect\ref{GFuncN}.\\

\subsection{The radiation reaction}

We calculate now the Lorentz force on the charges and the resulting radiation
resistance power for comparing with the radiated power. We will need the electric
field on charge $n$ at time $t$ due to charge $m$ at its retarded position
at the earlier time $t'$, so that:

\begin{equation}
|\mathbf{r'}_n(t)-\mathbf{r'}_m(t')|=c(t-t').
\label{light_cone}
\end{equation}

Using the expression for $\mathbf{r'}_k$ in eq.~(\protect\ref{rk}) we obtain

\begin{equation}
4d^2\sin^2\left[\frac{\omega(t-t')+2\pi(n-m)/N}{2}\right] =c^2(t-t')^2,
\label{light_cone0_5}
\end{equation}

which is exact for any $m$ and $n$. By definition, $t-t'>0$, so to take
the correct square root from the left side of this equation, we need to
know the connection between $m$ and $n$. To simplify, we restrict:

\begin{equation}
0\le m<n\le N-1,
\label{m_n_restrict}
\end{equation}

for which we obtain

\begin{equation}
2d\sin\Phi_{nm} =c(t-t'),
\label{light_cone1}
\end{equation}

where $\Phi_{nm}$ is defined by

\begin{equation}
\Phi_{nm}\equiv\frac{\omega(t-t')+2\pi(n-m)/N}{2}.
\label{PHI_nm}
\end{equation}

Now we isolate $t-t'$ from eq.~(\protect\ref{PHI_nm}) and set it in
eq.~(\protect\ref{light_cone1}), obtaining

\begin{equation}
\Phi_{nm}= \pi(n-m)/N +\beta\sin\Phi_{nm},
\label{light_cone2}
\end{equation}

which is an implicit equation, that can be solved by setting a 1st guess
$\Phi_{nm}= \pi(n-m)/N$ in the right side of the equation and recalculate
$\Phi_{nm}$ till convergence is obtained.

\begin{figure}[htbp]
\includegraphics[width=18cm]{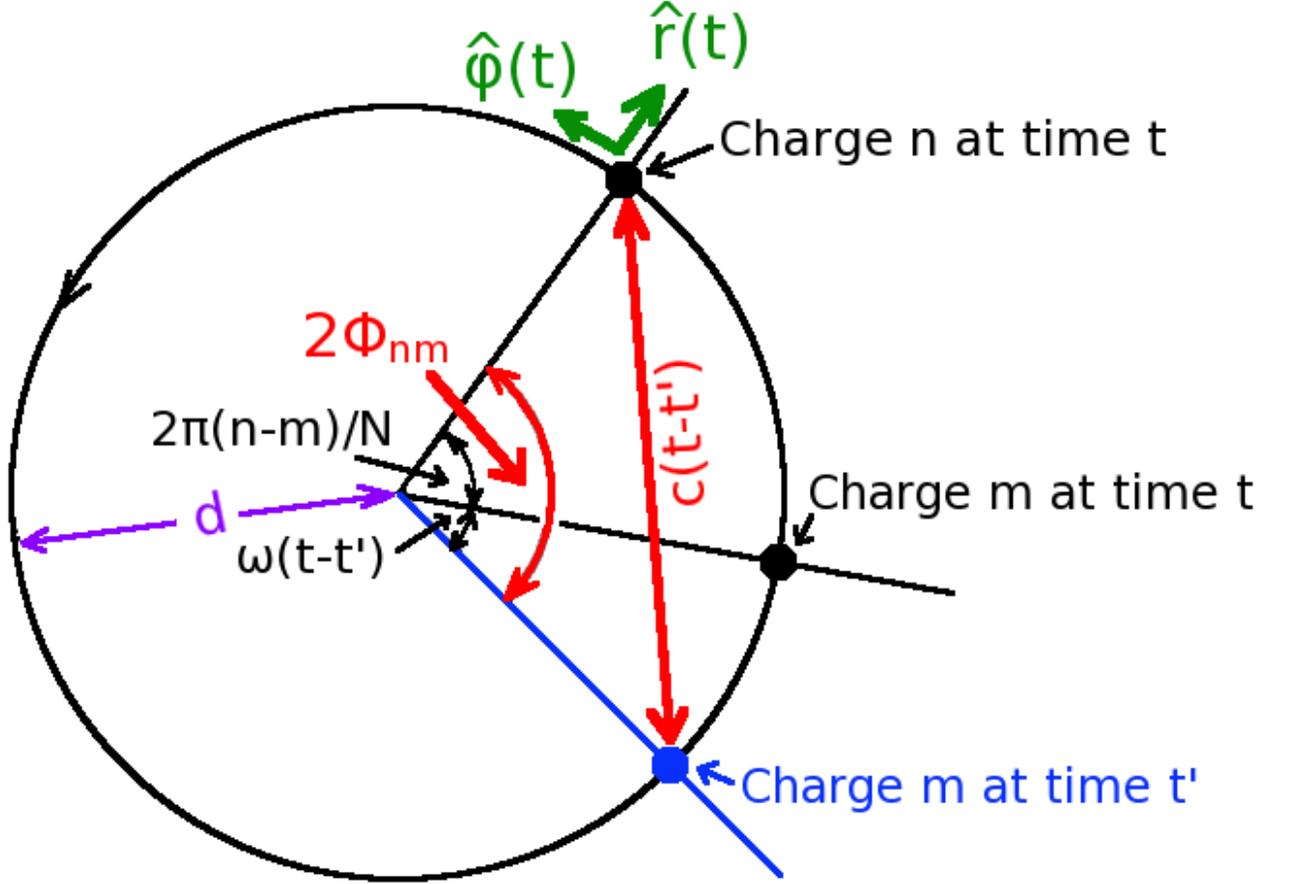}
\caption{Geometrical interpretation for
eqs.~(\protect\ref{light_cone1})-(\protect\ref{light_cone2}). The charges
rotate on the big circle of radius $d$. The retarded distance $c(t-t')$
is the big (red) segment on the $2\Phi_{mn}$ arc, hence equal to
$2d\sin\Phi_{mn}$ according to eq.~(\protect\ref{light_cone1}). We
also see that the big angle $2\Phi_{mn}$ (marked in red) equals the sum
of the angles $\omega(t-t')$ and $2\pi(n-m)/N$ according to
eq.~(\protect\ref{PHI_nm}). Two orthogonal unit vectors (green)
$\mathbf{\widehat{r}}(t)$ and $\boldsymbol{\widehat{\varphi}}(t)$
are drawn near charge $n$, representing the radial and tangential
directions of the moving charge.}
\label{lightcone_fig}
\end{figure}

Figure~(\protect\ref{lightcone_fig}) gives the geometrical interpretation
of eqs.~(\protect\ref{light_cone1})-(\protect\ref{light_cone2}).

Let us look at a simple example of solution for eq.~(\protect\ref{light_cone2}).
Say there are 4 charges, so their locations at $t=0$ are: $0^o$, $90^o$,
$180^o$ and $270^o$ for charges 0, 1, 2, 3 respectively and let us take
$\beta=0.7$. For calculating the effect of charges 0, 1 and 2 on charge 3,
we need to calculate $\Phi_{30}$, $\Phi_{31}$ and $\Phi_{32}$, which come
out: 2.67259, 2.15479 and 1.48268 respectively, in radians. The retarded
angle of charge $m$ is $270^o-2\Phi_{3m}$ - see
Figure~(\protect\ref{lightcone_fig}). So translating into degrees, we
get the retarded angles of $-36.256^o$, $23.079^o$ and $100.097^o$ for
charges 0, 1 and 2 respectively, which are all smaller than the current
angles of those charges.

It is to be mentioned that the solution $\Phi_{mn}$ of
eq.~(\protect\ref{light_cone2}) is time independent, meaning that
the angle difference between the current position of charge $n$ and
retarded position of charge $m$ does not depend on time. Because
the charges rotate, the only thing which depends on time are the
local unit vectors comoving with the charge $\mathbf{\widehat{r}}(t)$
and $\boldsymbol{\widehat{\varphi}}(t)$ - see
Figure~(\protect\ref{lightcone_fig}).

Now the electric field on charge $n$ due to charge $m$ at its retarded position
is given by \protect\cite{ianc_hor,rhorlich}

\begin{equation}
\mathbf{E}_{mn}=\frac{q/N}{4\pi\epsilon_0} \left[\frac{\mathbf{\widehat{R}}_{mn}-\boldsymbol{\beta}_m}{ \gamma^2 R_{mn}^2 (1-\boldsymbol{\beta}_m\cdot\mathbf{\widehat{R}}_{mn})^3} +
\frac{\mathbf{\widehat{R}}_{mn}\times[(\mathbf{\widehat{R}}_{mn}-\boldsymbol{\beta}_m)\times\boldsymbol{\dot{\beta}}_m]} {c R_{mn} (1-\boldsymbol{\beta}\cdot\mathbf{\widehat{R}}_{mn})^3}\right],
\label{E_mn}
\end{equation}

where the 2nd part is the far field which we used in eq.~(\protect\ref{Ek})
(after replacing $\mathbf{a}$ by $\boldsymbol{\dot{\beta}c}$ and
$1/\sqrt{\epsilon_0\mu_0}$ by $c$) and the first part is the near field
which behaves like $1/R^2$.

The quantities appearing in eq.~(\protect\ref{E_mn}) are:
$\boldsymbol{\beta}_m$ is the retarded velocity of charge $m$ (relative to $c$),
$R_{mn}$ is the distance between the retarded position of charge $m$ and the current
position of charge $n$ (and equals to $c(t-t')$ - see Figure~(\protect\ref{lightcone_fig})),
and $\mathbf{\widehat{R}}_{mn}$ is the unit vector pointing
from the retarded position of charge $m$ to the current position of charge $n$.

We need the field at charge $n$ in local components
$\mathbf{\widehat{r}}(t)$ and $\boldsymbol{\widehat{\varphi}}(t)$ (see
Figure~(\protect\ref{lightcone_fig})) and we calculate now all the needed
quantities to express the electric field $\mathbf{E}_{mn}$. So we obtain

\begin{equation}
\mathbf{\widehat{R}}_{mn}-\boldsymbol{\beta}_m=
\mathbf{\widehat{r}}(t)[\sin\Phi_{nm}-\beta\sin(2\Phi_{nm})]+
\boldsymbol{\widehat{\varphi}}(t)[\cos\Phi_{nm}-\beta\cos(2\Phi_{nm})]
\label{R_minus_beta}
\end{equation}

\begin{equation}
\mathbf{\widehat{R}}_{mn}\times[(\mathbf{\widehat{R}}_{mn}-\boldsymbol{\beta}_m)\times\boldsymbol{\dot{\beta}}_m]=
\omega\beta(\cos\Phi_{nm}-\beta)
[\mathbf{\widehat{r}(t)}\cos\Phi_{nm}-\boldsymbol{\widehat{\varphi}}(t)\sin\Phi_{nm}]
\label{R_cross_R_minus_beta_cross_dotbeta}
\end{equation}

\begin{equation}
\boldsymbol{\beta}_m\cdot\mathbf{\widehat{R}}_{mn}=\beta\cos\Phi_{nm}
\label{beta_dot_Rmn}
\end{equation}

and

\begin{equation}
R_{mn}=2d\sin\Phi_{nm}
\label{Rmn}
\end{equation}

which is easily derived also from Figure~(\protect\ref{lightcone_fig}), because
$R_{mn}=c(t-t')$, which equals to $2d\sin\Phi_{nm}$ according to
eq.~(\protect\ref{light_cone1}). Putting
eqs.~(\protect\ref{R_minus_beta})-(\protect\ref{Rmn}) in eq.~(\protect\ref{E_mn})
we obtain

\begin{align}
\mathbf{E}_{mn}=\frac{q/N}{4\pi\epsilon_0} &
\frac{1}{4 d^2\sin\Phi_{nm}(1-\beta\cos\Phi_{nm})^3} \notag\\
&\left[\frac{\mathbf{\widehat{r}}(t)[\sin\Phi_{nm}-\beta\sin(2\Phi_{nm})]+\boldsymbol{\widehat{\varphi}}(t)[\cos\Phi_{nm}-\beta\cos(2\Phi_{nm})]}{\gamma^2\sin\Phi_{nm}}+ \right. \notag\\
&\left. 2\beta^2(\cos\Phi_{nm}-\beta)[\mathbf{\widehat{r}(t)}\cos\Phi_{nm}-\boldsymbol{\widehat{\varphi}}(t)\sin\Phi_{nm}] \right]
\label{E_mn_1}
\end{align}

With the aid of this field we will calculate the total force acted on a charge
by {\it the other charges}, but we also need the ``self'' radiation reaction
force. For the most general case, this is given by \protect\cite{ianc_hor,rhorlich}

\begin{equation}
\mathbf{F}_{\text{self}}=\frac{(q/N)^2}{4\pi\epsilon_0 c^2}
\left[\frac{2}{3}\gamma^2\boldsymbol{\ddot{\beta}}+
2\gamma^4(\boldsymbol{\beta}\cdot\boldsymbol{\dot{\beta}})\boldsymbol{\dot{\beta}}+
\frac{2}{3}\gamma^4(\boldsymbol{\beta}\cdot\boldsymbol{\ddot{\beta}})\boldsymbol{\beta}+
2\gamma^6(\boldsymbol{\beta}\cdot\boldsymbol{\dot{\beta}})^2\boldsymbol{\beta}
\right].
\label{F_self}
\end{equation}

In our case, of circular motion, the velocity is perpendicular to the acceleration
so $\boldsymbol{\beta}\cdot\boldsymbol{\dot{\beta}}=0$, we therefore remain with 2
terms

\begin{equation}
\mathbf{F}_{\text{self}}=\frac{(q/N)^2}{4\pi\epsilon_0 c^2}
\left[\frac{2}{3}\gamma^2\boldsymbol{\ddot{\beta}}+
\frac{2}{3}\gamma^4(\boldsymbol{\beta}\cdot\boldsymbol{\ddot{\beta}})\boldsymbol{\beta}
\right].
\label{F_self_1}
\end{equation}

For the circular motion we know that
$\boldsymbol{\ddot{\beta}}=-\omega^2\boldsymbol{\beta}$, and using
$\omega=v/d=\beta c/d$, eq.~(\protect\ref{F_self_1}) reduces to

\begin{equation}
\mathbf{F}_{\text{self}}=-\frac{(q/N)^2\gamma^4}{6\pi\epsilon_0 d^2} \beta^2 \boldsymbol{\beta},
\label{F_self_2}
\end{equation}

showing that the self reaction force is in the direction opposite to the velocity
of the charge.

Now we chose $n$ to be the ``last'' charge, i.e. $n=N-1$, hence the restriction
in eq.~(\protect\ref{m_n_restrict}) holds, and calculate the total force on it,
given by the self force plus the force acted by all other charges (i.e. the Lorentz
force):

\begin{equation}
\mathbf{F}_{N-1}=\mathbf{F}_{\text{self}}+(q/N)\sum_{m=0}^{N-2}[\mathbf{E}_{m,N-1}+\mathbf{v}_{N-1}(t)\times\mathbf{B}_{m,N-1}],
\label{F_N_minus1}
\end{equation}

where $\mathbf{v}_{N-1}(t)$ is the velocity of charge $N-1$ and $\mathbf{B}_{m,N-1}$
is the retarded magnetic field on charge $N-1$ due to charge $m$.

The dumping power on charge $N-1$ is given by
$\mathbf{v}_{N-1}(t)\cdot\mathbf{F}_{N-1}$, therefore we do not need the magnetic
part, obtaining

\begin{equation}
P_{\text{dump 1 charge}}=\mathbf{v}_{N-1}(t)\cdot\mathbf{F}_{N-1}=
\mathbf{v}_{N-1}(t)\cdot\mathbf{F}_{\text{self}}+
(q/N)\sum_{m=0}^{N-2}\mathbf{v}_{N-1}(t)\cdot\mathbf{E}_{m,N-1},
\label{P_dump_N_1}
\end{equation}

The velocity of the given charge $N-1$ is in the tangential direction,
i.e. $\mathbf{v}_{N-1}(t)=c\beta\boldsymbol{\widehat{\varphi}}(t)$,
so only the tangential part of eq.~(\protect\ref{E_mn_1}) affects
the power. We obtain

\begin{align}
P_{\text{dump 1 charge}}=c\beta & \left\{-\frac{(q/N)^2\gamma^4}{6\pi\epsilon_0 d^2} \beta^3+ \right.
\sum_{m=0}^{N-2} \frac{(q/N)^2}{4\pi\epsilon_0}
\frac{1}{4 d^2\sin\Phi_{N-1,m}(1-\beta\cos\Phi_{N-1,m})^3}\notag \\
&\left[\frac{\cos\Phi_{N-1,m}-\beta\cos(2\Phi_{N-1,m})}{\gamma^2\sin\Phi_{N-1,m}}
- \left. 2\beta^2(\cos\Phi_{N-1,m}-\beta)\sin\Phi_{N-1,m}\right]\right\}.
\label{P_dump_N_1_1}
\end{align}

Because the dumping power on one charge does not depend on time and
by symmetry is the same for all charges, the total dumping power
on the whole system of charges is just N times the above:

\begin{align}
P_{\text{dump}}=c\beta & \left\{-\frac{q^2\gamma^4}{6\pi\epsilon_0 N d^2} \beta^3+ \right.
\sum_{m=0}^{N-2} \frac{q^2}{4\pi\epsilon_0 N}
\frac{1}{4 d^2\sin\Phi_{N-1,m}(1-\beta\cos\Phi_{N-1,m})^3}\notag \\
&\left[\frac{\cos\Phi_{N-1,m}-\beta\cos(2\Phi_{N-1,m})}{\gamma^2\sin\Phi_{N-1,m}}
- \left. 2\beta^2(\cos\Phi_{N-1,m}-\beta)\sin\Phi_{N-1,m}\right]\right\}.
\label{P_dump}
\end{align}

The dumping power must be identical with
the radiated power, with a minus sign, so to compare them, we
may factor out $-\left.P\right|_{\,N=1}$ from eq.~(\protect\ref{larmor2}):

\begin{equation}
P_{\text{dump}}=-\left.P\right|_{\,N=1}G_{\text{dump}}(\beta,N),
\label{P_dump_1}
\end{equation}

where

\begin{align}
G_{\text{dump}}(\beta,N)=
 & \frac{1}{N}-
\frac{3}{8N\beta^3\gamma^4}\sum_{m=0}^{N-2}
\frac{1}{\sin\Phi_{N-1,m}(1-\beta\cos\Phi_{N-1,m})^3}\notag \\
&\left[\frac{\cos\Phi_{N-1,m}-\beta\cos(2\Phi_{N-1,m})}{\gamma^2\sin\Phi_{N-1,m}}
-2\beta^2(\cos\Phi_{N-1,m}-\beta)\sin\Phi_{N-1,m}\right].
\label{G_dump_1}
\end{align}

which clearly shows that for $N=1$, the sum is 0, so that
$G_{\text{dump}}(\beta,1)=1$ for any $\beta$. The numerical calculation of
eq.~(\protect\ref{G_dump_1}) shows identical results with those
of $G$ calculated from eq.~(\protect\ref{G5}), as shown in
Figure~\protect\ref{GFuncN}.

\section{Asymptotic result for many charges}

To understand how the radiation goes to 0 when the number of
charges $N$ goes to infinity, one may try to approximate either
$G$ from eq.~(\protect\ref{G5}) or $G_{\text{dump}}$ from
eq.~(\protect\ref{G_dump_1}), for big $N$. Although $G_{\text{dump}}$
seems more compact, it is more difficult to handle, and we shall
develop $G$ for large $N$.

As mentioned in the previous section (see
Figures~\protect\ref{sin_cos_01_3}-\protect\ref{sin_cos_05_10}),
for large $N$ the functions Fc and Fs tend to be harmonic, hence we
may approximate them by the first term of their Fourier series.

For a function $\text{a}(x)$ with periodicity $X$, we specify the Fourier
coefficients by

\begin{equation}
A_n=\int_0^{X} \text{a}(x) e^{-i 2\pi n x/X} dx
\label{A_n}
\end{equation}

and $\text{a}(x)$ is represented by its Fourier series

\begin{equation}
\text{a}(x)=\frac{1}{X}\sum_{n=-\infty}^{\infty} A_n e^{i 2\pi n x/X} 
\label{a_x}
\end{equation}

For brevity, to refer to the functions Fc and Fs we call them
$\text{Fc,s}$ (as in eq.~(\protect\ref{Avfc})). We know those functions have a
periodicity of $2\pi/N$ in $\varphi'$ (see eq.~(\protect\ref{Fc_invariance})),
so we define their Fourier coefficients:

\begin{equation}
\text{Ac,s}_n=\int_0^{2\pi/N} \text{Fc,s}(\varphi') e^{-iNn\varphi'} d\varphi',
\label{Acs_n}
\end{equation}

and the functions Fc and Fs are expressed as:

\begin{equation}
\text{Fc,s}(\varphi')=\frac{1}{2\pi/N}\sum_{n=-\infty}^{\infty} \text{Ac,s}_n e^{iNn\varphi'}
\label{Fcs}
\end{equation}

Now we calculate the Fourier coefficients $\text{Ac,s}_n$ in
eq.~(\protect\ref{Acs_n}). The integrand is periodic in $2\pi/N$,
so increasing the integration interval to $2\pi$ multiplies the result
by $N$, hence we may express

\begin{equation}
\text{Ac,s}_n=\frac{1}{N}\int_0^{2\pi} \text{Fc,s}(\varphi') e^{-iNn\varphi'} d\varphi',
\label{Acs_n1}
\end{equation}

and by using the definitions of \text{Fc,s} (eqs.~(\protect\ref{Fc}) and (\protect\ref{Fs}))
and the property of $\phi_k$ from eq.~(\protect\ref{phik_m}), we get

\begin{equation}
\text{Ac,s}_n=\frac{1}{N}\int_0^{2\pi}\sum_{k=0}^{N-1}\text{fc,s}(\phi_0(\varphi'+2\pi k/N)) e^{-iNn\varphi'} d\varphi'.
\label{Acs_n2}
\end{equation}

We interchange the sum and the integral and change variable
$\varphi''=\varphi'+2\pi k/N$, obtaining

\begin{equation}
\text{Ac,s}_n=\frac{1}{N}\sum_{k=0}^{N-1}e^{-i n 2\pi k}\int_{2\pi k/N}^{2\pi(1+k/N)}\text{fc,s}(\phi_0(\varphi'')) e^{-iNn\varphi''} d\varphi''.
\label{Acs_n3}
\end{equation}

In the above integral, $\text{fc,s}$ has a periodicity of $2\pi$ and
$e^{-iNn\varphi''}$ has a periodicity of $2\pi/N$, therefore the
integrand is periodic by $2\pi$. We may therefore move the integration
range to be between 0 and $2\pi$, showing that the integral does not
depend on $k$. After renaming $\varphi''$ to $\varphi'$ we obtain

\begin{equation}
\text{Ac,s}_n=\frac{1}{N}\left(\sum_{k=0}^{N-1}e^{-i n 2\pi k}\right)\int_{0}^{2\pi}\text{fc,s}(\phi_0(\varphi')) e^{-iNn\varphi'} d\varphi'=\int_{0}^{2\pi}\text{fc,s}(\phi_0(\varphi')) e^{-iNn\varphi'} d\varphi',
\label{Acs_n4}
\end{equation}

because $e^{-i n 2\pi k}=1$ for any $k$. We see that the $n$ Fourier
coefficient of $\text{Fc,s}$ is actually the $Nn$ Fourier coefficient
of $\text{fc,s}(\phi_0)$, which have a $2\pi$ periodicity. This means
that if we represented the $\text{fc,s}$ functions by their Fourier
components and evaluated $\text{Fc,s}=\sum_{k=0}^{N-1}\text{fc,s}(\phi_k)$,
all Fourier components would cancel out except of the $nN$ components,
i.e. the 0, $N$, $2N$, etc.
The 0 Fourier coefficient is 0, because $\text{fc,s}$ have 0 DC level (see 
eqs~(\protect\ref{Avfc2}-\protect\ref{Avfc4})), and clearly the second
Fourier coefficient of $\text{Fc,s}$, which is the $2N$ Fourier
coefficient of $\text{fc,s}$, is much smaller than the first coefficient
for large $N$, as evident also from
Figures~(\protect\ref{sin_cos_01_3})-(\protect\ref{sin_cos_05_10}).

Therefore, for large $N$ we get the asymptotic $\text{Fc,s}$ from its first
Fourier coefficient (i.e. coefficients $1$ and $-1$), so that we get
from eq.~(\protect\ref{Fcs}):

\begin{equation}
\text{Fc,s}(\varphi')\rightarrow\frac{N}{2\pi}
\left(\text{Ac,s}_1 e^{iN 1 \varphi'}+\text{Ac,s}_{-1} e^{iN(-1)\varphi'}\right).
\label{Fcs_1}
\end{equation}

Because $\text{Fc,s}$ are real, $\text{Ac,s}_{-1}=\text{Ac,s}^*_1$ and
we obtain

\begin{equation}
\text{Fc,s}(\varphi')\rightarrow\frac{N}{\pi}
|\text{Ac,s}_1| \cos\left[N\varphi'+ \arg(\text{Ac,s}_1)\right].
\label{Fcs_1_2}
\end{equation}

So we have to calculate the first Fourier coefficients of the Fc and Fs
functions, $\text{Ac,s}_1$. From eq.~(\protect\ref{Acs_n4}) we get

\begin{equation}
\text{Ac,s}_1=\int_{0}^{2\pi}\text{fc,s}(\phi_0(\varphi')) e^{-iN\varphi'} d\varphi',
\label{Acs_1}
\end{equation}

We change variable to $\phi_0$ and according to eq.~(\protect\ref{dphi_over_dphik}) we
have $d\varphi'/d\phi_0=-1-p\hspace{1mm}\sin\phi_0$, so we obtain:

\begin{equation}
\text{Ac,s}_1=\int_{\phi_0(0)}^{\phi_0(0)-2\pi}\frac{\text{hc,s}(\phi_0)}{(1+p\sin\phi_0)^3} e^{-iN(-\phi_0+p\cos\phi_0)} d\phi_0 (-1-p\hspace{1mm}\sin\phi_0),
\label{Acs_1_1}
\end{equation}

where $\text{hc,s}$ is an abbreviation for the functions
$\text{hc}(\phi_0)$ and $\text{hs}(\phi_0)$, defined in
eqs.~(\protect\ref{hc}) and (\protect\ref{hs}), respectively.

The integrand being periodic on $2\pi$, we may shift the limits by any value
getting

\begin{equation}
\text{Ac,s}_1=\int_{0}^{2\pi}\frac{\text{hc,s}(\phi_0)}{(1+p\sin\phi_0)^2} e^{iN\phi_0} e^{-iNp\cos\phi_0} d\phi_0.
\label{Acs_1_2}
\end{equation}

We change variable $z=\exp(i\phi_0)$ to solve this integral on the complex plane,
obtaining:

\begin{equation}
\text{Ac,s}_1=\frac{2i}{p^2} \oint_{C}dz \frac{\text{lc,s}(z)}{(z^2+b^{-1}z-1)^2} z^N e^{Nb(z+z^{-1})}\equiv\frac{2i}{p^2} \oint_{C}dz\, f(z)
\label{Acs_1_c}
\end{equation}

where $b$ is the pure imaginary number defined in
eq.~(\protect\ref{b}), $C$ is the counterclockwise unit circle
integration contour (see Figure~(\protect\ref{poles})) and the
functions $\text{lc,s}$ are abbreviations for $\text{lc}(z)$ and
$\text{ls}(z)$ defined in eqs.~(\protect\ref{lc}) and
(\protect\ref{ls}), respectively.  To ease on further manipulations we
called the integrand $f(z)$.

The integrand has two 2nd order poles, $z_{1,2}$, defined in
eq.~(\protect\ref{z12}) and only $z_1$ lies inside the integration
contour $C$ - see Figure~(\protect\ref{poles}). In addition there is
an essential singularity at $z=0$, because of the $z^{-1}$ in the
exponent.

We first calculate the residue at $z=z1$. Rewriting $f(z)$

\begin{equation}
f(z)=\text{Lc,s}(z) z^N e^{Nb(z+z^{-1})},
\label{fz}
\end{equation}

where $\text{Lc,s}(z)$ is defined in eq.~(\protect\ref{Lz}), we obtain

\begin{equation}
\text{Res}(f,z1)=\lim_{z\rightarrow z_1}\frac{d}{dz}[f(z)(z-z_1)^2]=
\left.\frac{d}{dz}\left[\frac{\text{lc,s}(z)}{(z-z_2)^2}z^N e^{Nb(z+z^{-1})}\right]\right|_{z=z_1}
\label{Res_0}
\end{equation}

which evaluates to

\begin{equation}
\left.\text{Res}(f,z1)=\frac{\text{lc,s}'(z_1)(z_1-z_2)-2\text{lc,s}'(z_1)}{(z_1-z_2)^3} z_1^N e^{Nb(z_1+z_1^{-1})}+ \frac{\text{lc,s}(z_1)}{(z_1-z_2)^2} \frac{d}{dz}\left[z^N e^{Nb(z+z^{-1})}\right]\right|_{z=z_1}.
\label{Res_0_1}
\end{equation}

We already showed that the first part is 0 - (see eq.~(\protect\ref{Res_00})).
This is because the integral in eq.~(\protect\ref{Acs_1_2}) reduces for $N=0$
to the integral in eq.~(\protect\ref{Avfc2}) (up to $2\pi$). So
we are left with

\begin{equation}
\left.\text{Res}(f,z1)=\frac{\text{lc,s}(z_1)}{(z_1-z_2)^2} \frac{d}{dz}\left[z^N e^{Nb(z+z^{-1})}\right]\right|_{z=z_1}=\frac{\text{lc,s}(z_1)}{(z_1-z_2)^2} z_1^{N-1} e^{Nb(z_1+z_1^{-1})}N[1+b(z_1-z_1^{-1})].
\label{Res_0_2}
\end{equation}

By using $z_1=-1/z_2$ and $z_1+z_2 =-2i/p=-1/b$, we see that this part is 0
too, hence the contribution of the pole at $z=z_1$ is 0. We may therefore
exclude this pole from the integration contour, and rewrite eq.~(\protect\ref{Acs_1_c})

\begin{equation}
\text{Ac,s}_1=\frac{2i}{p^2} \oint_{C_1}dz \frac{\text{lc,s}(z)}{(z^2+b^{-1}z-1)^2} z^N e^{Nbz} e^{Nbz^{-1}}=
\frac{2i}{p^2} \sum_{n=0}^{\infty}\oint_{C_1}dz \frac{\text{lc,s}(z)}{(z^2+b^{-1}z-1)^2} e^{Nbz}\frac{(Nb)^n}{n! z^{n-N}},
\label{Acs_1_c1}
\end{equation}

where $C_1$ is the counterclockwise circle of radius smaller than
$|z_1|$, shown in Figure~(\protect\ref{poles}). Inside this
integration contour we have only the essential singularity at $z=0$,
and for handling it, we represented the exponent with negative powers
of $z$ as a Laurent series. The terms $n\le N$ are analytic inside $C_1$,
hence contribute 0 to the integral, so by changing the summation variable
$n'=n-(N+1)$ and remaining $n'$ to $n$ we obtain

\begin{equation}
\text{Ac,s}_1=\frac{2i}{p^2} \sum_{n=0}^{\infty}\frac{(Nb)^{n+N+1}}{(n+N+1)!}\oint_{C_1}dz \text{Lc,s}(z) e^{Nbz}\frac{1}{z^{n+1}}\equiv\frac{2i}{p^2} \sum_{n=0}^{\infty}\frac{(Nb)^{n+N+1}}{(n+N+1)!}\oint_{C_1}dz\hspace{1mm}g(z),
\label{Acs_1_c2}
\end{equation}

where we used again the definition of $\text{Lc,s}(z)$ from eq.~(\protect\ref{Lz}),
and the integrand has been called $g(z)$, to ease manipulations.
We calculate now the residue of $g(z)$

\begin{equation}
\left.\text{Res}(g,0)=\frac{1}{n!}\frac{d^n}{dz^n}[\text{Lc,s}(z) e^{Nbz}]\right|_{z=0}=
\frac{1}{n!}\sum_{m=0}^n \binom{n}{m} \left(\frac{d^m}{dz^m}\text{Lc,s}(z)\right)\left.\left(\frac{d^{n-m}}{dz^{n-m}}e^{Nbz}\right)\right|_{z=0},
\label{Res_g_0}
\end{equation}

which comes out

\begin{equation}
\text{Res}(g,0)=(Nb)^n\sum_{m=0}^n \frac{(Nb)^{-m}}{(n-m)!m!} \frac{d^m}{dz^m}\text{Lc,s}(z) |_{z=0},
\label{Res_g_0_1}
\end{equation}

We start with Lc. To handle this derivative we express it as

\begin{equation}
\left.\frac{d^m}{dz^m}\text{Lc}(z)\right|_{z=0}=\left. \frac{d^{m+1}}{dz^{m+1}}{\frac{z}{1-b^{-1}z-z^2}}\right|_{z=0}
\label{dm_Lc}
\end{equation}

The last rational function is the Fibonacci polynomials generating function,
with argument $b^{-1}$. Hence the result is $(m+1)!$ multiplied by the $m+1$
Fibonacci polynomial. This may be directly calculated by factorizing the
Fibonacci generating function to obtain

\begin{equation}
\left.\frac{d^m}{dz^m}\text{Lc}(z)\right|_{z=0}=
\frac{1}{\sqrt{b^{-2}+4}} \left.\frac{d^{m+1}}{dz^{m+1}}\left[\frac{1}{1-z_1^{-1}z}-\frac{1}{1-z_2^{-1}z}\right]\right|_{z=0}
\label{dm_Lc_1}
\end{equation}

which results in

\begin{equation}
\left.\frac{d^m}{dz^m}\text{Lc}(z)\right|_{z=0}=
(m+1)!\frac{\left(z_1^{-1}\right)^{m+1}-\left(z_2^{-1}\right)^{m+1}}{\sqrt{b^{-2}+4}}
\label{dm_Lc_2}
\end{equation}

or explicitly

\begin{equation}
\left.\frac{d^m}{dz^m}\text{Lc}(z)\right|_{z=0}=
(m+1)!\frac{\left(b^{-1}+\sqrt{b^{-2}+4}\right)^{m+1}-\left(b^{-1}-\sqrt{b^{-2}+4}\right)^{m+1}}{2^{m+1}\sqrt{b^{-2}+4}}
\label{dm_Lc_3}
\end{equation}

The Ls case is handled similarly. We express:

\begin{equation}
\left.\frac{d^m}{dz^m}\text{Ls}(z)\right|_{z=0}=ib \left. \frac{d^{m+1}}{dz^{m+1}}{\frac{2-b^{-1}z}{1-b^{-1}z-z^2}}\right|_{z=0}
\label{dm_Ls}
\end{equation}

The last rational function is the Lucas polynomials generating function,
with argument $b^{-1}$. Hence the result is $(m+1)!$ multiplied by the $m+1$
Lucas polynomial. This may be directly calculated by factorizing the
Lucas generating function to obtain

\begin{equation}
\left.\frac{d^m}{dz^m}\text{Ls}(z)\right|_{z=0}=
ib \left.\frac{d^{m+1}}{dz^{m+1}}\left[\frac{1}{1-z_1^{-1}z}+\frac{1}{1-z_2^{-1}z}\right]\right|_{z=0}
\label{dm_Ls_1}
\end{equation}

which results in

\begin{equation}
\left.\frac{d^m}{dz^m}\text{Ls}(z)\right|_{z=0}=
ib(m+1)!\left[\left(z_1^{-1}\right)^{m+1}+\left(z_2^{-1}\right)^{m+1}\right]
\label{dm_Ls_2}
\end{equation}

or explicitly

\begin{equation}
\left.\frac{d^m}{dz^m}\text{Ls}(z)\right|_{z=0}=
ib(m+1)!\frac{\left(b^{-1}+\sqrt{b^{-2}+4}\right)^{m+1}+\left(b^{-1}-\sqrt{b^{-2}+4}\right)^{m+1}}{2^{m+1}}
\label{dm_Ls_3}
\end{equation}

By using eqs.~(\protect\ref{dm_Lc_3}), (\protect\ref{Res_g_0_1}) and
(\protect\ref{Acs_1_c2}) and replacing $b=-0.5ip$ we obtain a closed form expression
for $\text{Ac}_1$

\begin{align}
\text{Ac}_1=\frac{-2\pi(-i)^{N+1}}{\sqrt{1-p^2}} & \sum_{n=0}^{\infty}\frac{(-1)^n (N/2)^{2n+N+1}}{(n+N+1)!}  \sum_{m=0}^n \frac{(-1)^m (m+1)p^{2(n-m)+N-1}}{(N/2)^m(n-m)!} \notag\\
 & \left[\left(1+\sqrt{1-p^2}\right)^{m+1}-\left(1-\sqrt{1-p^2}\right)^{m+1}\right]
\label{Ac1}
\end{align}

and for $\text{As}_1$

\begin{align}
\text{As}_1=-2\pi(-i)^{N+1}i & \sum_{n=0}^{\infty}\frac{(-1)^n (N/2)^{2n+N+1}}{(n+N+1)!}  \sum_{m=0}^n \frac{(-1)^m (m+1)p^{2(n-m)+N-1}}{(N/2)^m(n-m)!} \notag\\
& \left[\left(1+\sqrt{1-p^2}\right)^{m+1}+\left(1-\sqrt{1-p^2}\right)^{m+1}\right]
\label{As1}
\end{align}

We remark that if $N$ is a multiple of 4, the angle of $\text{Ac}_1$ is $90^o$, and
each increment of $N$ removes $90^o$ from the angle of $\text{Ac}_1$. Also we see that
the angle of $\text{As}_1$ is always bigger by $90^o$ than the angle of $\text{Ac}_1$
and this is visible in Figures~(\protect\ref{sin_cos_01_3}), (\protect\ref{sin_cos_03_3})
and (\protect\ref{sin_cos_05_10}). We shall name those angles $\varphi_c$ and $\varphi_s$
in the following calculations.

So by setting eqs.~(\protect\ref{Fcs_1_2}) in (\protect\ref{G5}), and inverting the
integration order between $dp$ and $d\varphi'$ we obtain:

\begin{align}
G(\beta,N\rightarrow\infty)=\frac{3}{4\pi N\gamma^4\beta^2} & \left[\int_0^{\beta} \frac{dp\, p \left(\frac{N}{\pi}|\text{As}_1|\right)^2 }{\sqrt{1-(p/\beta)^2}} \int_0^{2\pi/N}d\varphi' \right. \cos^2(N\varphi'+\varphi_s) +  \notag\\
& \left. \int_0^{\beta} dp\, p \left(\frac{N}{\pi}|\text{Ac}_1|\right)^2 \sqrt{1-(p/\beta)^2} \int_0^{2\pi/N}d\varphi' \cos^2(N\varphi'+\varphi_c)\right].
\label{G5_bigN}
\end{align}

The $d\varphi'$ integrals result in half the integration interval, i.e. $\pi/N$, so
after simplifying we obtain

\begin{equation}
G(\beta,N\rightarrow\infty)=\frac{3}{4\pi N\gamma^4\beta^2} \frac{N}{\pi} \int_0^{\beta} dp\, p
\left(\frac{|\text{As}_1|^2 }{\sqrt{1-(p/\beta)^2}} + |\text{Ac}_1|^2 \sqrt{1-(p/\beta)^2}\right).
\label{G5_bigN_1}
\end{equation}

\begin{figure}[htbp]
\includegraphics[width=15cm]{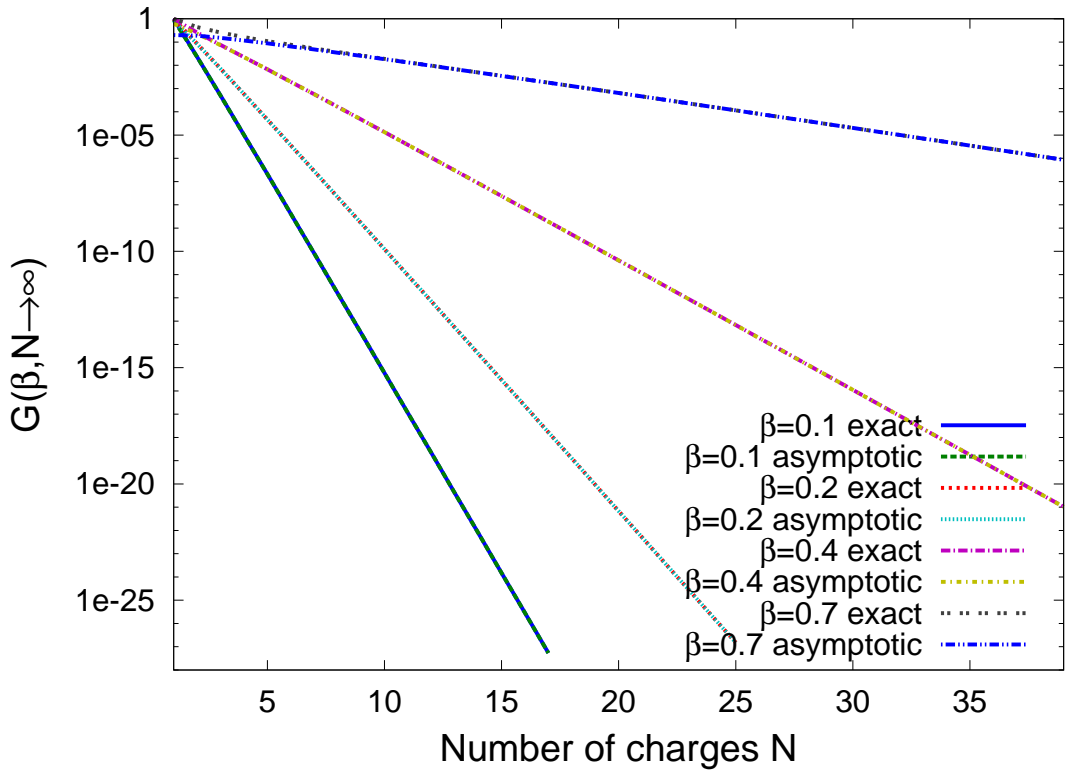}
\caption{Asymptotic results for $G$ according to eq.~(\protect\ref{G5_bigN_1}) versus
exact results according to eq.~(\protect\ref{G5}) as function of the number
of charges $N$, for different values of $\beta$. The asymptotic approximation is
accurate even for a few number of charges. As $\beta$ gets bigger, the difference
between asymptotic and exact is more visible for small $N$.}
\label{asymptoticGFuncN_infty}
\end{figure}

Figure~(\protect\ref{asymptoticGFuncN_infty}) shows the asymptotic results of $G$ for
large $N$ according to eqs.~(\protect\ref{G5_bigN_1}), (\protect\ref{Ac1}) and
(\protect\ref{As1}), compared with the exact result from eq.~(\protect\ref{G5}).

The asymptotic result is much easier calculable than the exact one, and does not require 
to solve for each step the implicit equation (\protect\ref{phik2}), but is still not given
by a simple formula.

We will calculate in the next subsection an asymptotic expression
for small $\beta$, i.e. $G(\beta\rightarrow 0,N\rightarrow\infty)$, and for
this case one arrives to a simple formula, as we shall see below.

\subsection{Asymptotic result for many charges and low velocity}

For small $\beta$, $|p|\ll 1$, hence
$\sqrt{1-p^2}$ in the denominator of eq.~(\protect\ref{Ac1}) may be set to 1,
and $1+\sqrt{1-p^2}\approx 2$, neglecting $1-\sqrt{1-p^2}$ in
eqs.~(\protect\ref{Ac1}) and (\protect\ref{As1}). So we obtain

\begin{equation}
\text{Ac}_1|_{|p|\rightarrow 0}=-2\pi(-i)^{N+1}\sum_{n=0}^{\infty}\frac{(-1)^n (N/2)^{2n+N+1}}{(n+N+1)!}  \sum_{m=0}^n \frac{(-1)^m (m+1)p^{2(n-m)+N-1}}{(N/2)^m(n-m)!}2^{m+1}
\label{Ac1_p=0}
\end{equation}

and

\begin{equation}
\text{As}_1|_{|p|\rightarrow 0}=i\text{Ac}_1|_{|p|\rightarrow 0}.
\label{As1_p=0}
\end{equation}

Rearranging eq.~(\protect\ref{Ac1_p=0}) we obtain

\begin{equation}
\text{Ac}_1|_{|p|\rightarrow 0}=-4\pi(-i)^{N+1}p^{N-1}\sum_{n=0}^{\infty}\frac{(-1)^n (N/2)^{2n+N+1}p^{2n}}{(n+N+1)!} \sum_{m=0}^n \frac{(m+1)}{(n-m)!}\left(\frac{-4}{Np^2}\right)^{m},
\label{Ac1_p=0_1}
\end{equation}

and defining $K\equiv -4/(Np^2)$, one may express the sum over $m$ in
eq.~(\protect\ref{Ac1_p=0_1}) as a derivative with respect to $K$ as
follows

\begin{equation}
\partial_K\sum_{m=0}^n \frac{K^{m+1}}{(n-m)!},
\label{sum_m}
\end{equation}

and by changing the summation variable $m'=n-m$, this is written as

\begin{equation}
\partial_K\sum_{m'=n}^0 \frac{K^{n-m'+1}}{m'!}=
\partial_K\left[K^{n+1}\sum_{m'=0}^n \frac{(1/K)^{m'}}{m'!}\right].
\label{sum_mp}
\end{equation}

We may sum exactly the last sum over $m'$ to obtain

\begin{equation}
\sum_{m'=0}^n \frac{(1/K)^{m'}}{m'!}=\exp(1/K)\frac{\Gamma(n+1,1/K)}{n!}\approx\exp(1/K),
\label{partial_exp}
\end{equation}

where $\Gamma$ with 2 arguments is the incomplete gamma function. We are interested
in small $|p|$, so for $|1/K|=Np^2/4\ll 1$, we obtain the approximated result given
in eq.~(\protect\ref{partial_exp}), which means that for small argument, the
exponential series in eq.~(\protect\ref{partial_exp}) needs
very few terms to converge to an exponent. So continuing the calculation started
in eq.~(\protect\ref{sum_mp}) we obtain

\begin{equation}
\partial_K\left[K^{n+1}\exp(1/K)]\right]=K^n\exp(1/K)[n+1-1/K]\approx(n+1)K^n=
(n+1)\left(\frac{-4}{Np^2}\right)^n,
\label{sum_mp_1}
\end{equation}

where the last approximation used again the fact that $|1/K|\ll 1$. Now using the
result from eq.~(\protect\ref{sum_mp_1}) in eq.~(\protect\ref{Ac1_p=0_1}), we
obtain

\begin{equation}
\text{Ac}_1|_{|p|\rightarrow 0}=-4\pi(-i)^{N+1}p^{N-1}(N/2)^{N+1}
\sum_{n=0}^{\infty}\frac{n+1}{(n+N+1)!} N^n.
\label{Ac1_p=0_2}
\end{equation}

The above may be summed exactly, obtaining

\begin{equation}
\sum_{n=0}^{\infty}\frac{n+1}{(n+N+1)!} N^n=
\frac{N^{N+3}\Gamma(N) - e^N \left(\Gamma(N+2)\Gamma(N+1,N) - N\Gamma(N)\Gamma(N+2,N)\right) }{N^{(2+N)}\Gamma(N)\Gamma(N+2)}.
\label{sum_n}
\end{equation}

For large $N$, $\Gamma(N+1,N)\approx\frac{1}{2}\Gamma(N+1)$ and 
$\Gamma(N+2,N)\approx\frac{1}{2}\Gamma(N+2)$, hence the expression
multiplying the exponent in eq.~(\protect\ref{sum_n}) tents to 0,
remaining with

\begin{equation}
\sum_{n=0}^{\infty}\frac{n+1}{(n+N+1)!} N^n\approx
\frac{N^{-(2+N)}\left[N^{N+3}\Gamma(N)\right] }{\Gamma(N)\Gamma(N+2)}=\frac{N}{\Gamma(N+2)}\approx\frac{1}{N!}.
\label{sum_n_1}
\end{equation}

Putting eq.~(\protect\ref{sum_n_1}) in (\protect\ref{Ac1_p=0_2}) we obtain

\begin{equation}
\text{Ac}_1|_{|p|\rightarrow 0}=-4\pi(-i)^{N+1}p^{N-1}(N/2)^{N+1}\frac{1}{N!}\approx
(-i)^{N+1}\sqrt{2\pi N}(e/2)^N p^{N-1},
\label{Ac1_p=0_3}
\end{equation}

where the last expression has been obtained by using the Stirling approximation, and
$\text{As}_1|_{|p|\rightarrow 0}=i\text{Ac}_1|_{|p|\rightarrow 0}$ according to
eq.~(\protect\ref{As1_p=0}).

Now we may perform the integral in eq.~(\protect\ref{G5_bigN_1}), which becomes for
small $\beta$

\begin{equation}
G(\beta\rightarrow 0,N\rightarrow\infty)=\frac{3\times 2\pi N (e/2)^{2N}}{4\pi^2 \gamma^4\beta^2} \int_0^{\beta} dp\, p^{2N-1}
\left(\frac{1}{\sqrt{1-(p/\beta)^2}} + \sqrt{1-(p/\beta)^2}\right).
\label{G5_bigN_2}
\end{equation}

Because $\beta\rightarrow 0$, we may set $\gamma=1$, and we obtain

\begin{equation}
G(\beta\rightarrow 0,N\rightarrow\infty)=\frac{3 N (e/2)^{2N}}{2\pi \beta^2}
\frac{\beta^{2N}(N+1)\sqrt{\pi}\,\Gamma(N)}{2\Gamma(N+3/2)}.
\label{G5_bigN_3}
\end{equation}

For large $N$, $\Gamma(N+3/2)/\Gamma(N)\approx N^{3/2}$, so we obtain

\begin{equation}
G(\beta\rightarrow 0,N\rightarrow\infty)=\frac{3 N (e/2)^{2N}}{2\pi \beta^2}
\frac{\beta^{2N}(N+1)\sqrt{\pi}}{2N^{3/2}}\approx
\frac{3 (e\beta/2)^{2N}\sqrt{N}}{4\sqrt{\pi} \beta^2}.
\label{G5_bigN_4}
\end{equation}

\begin{figure}[htbp]
\includegraphics[width=15cm]{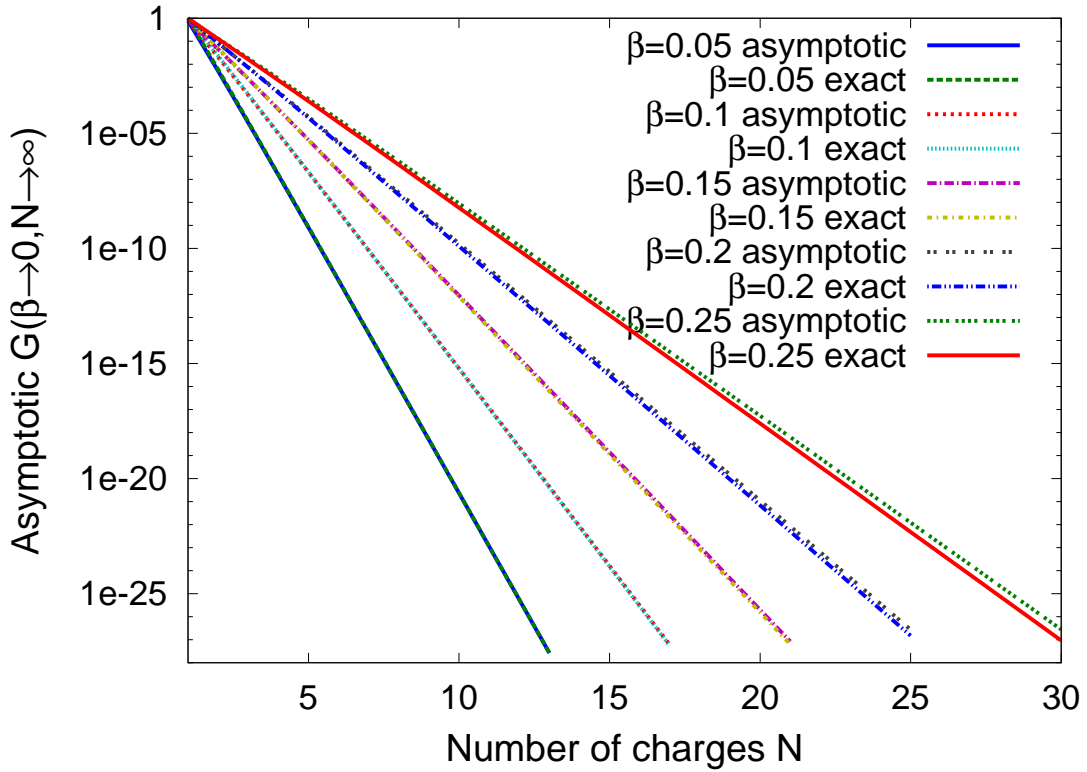}
\caption{Asymptotic results for $G$ according to eq.~(\protect\ref{G5_bigN_3}) versus
exact results according to eq.~(\protect\ref{G5}) as function of the number
of charges $N$, for different values of $\beta$. For $\beta=0.05$ the asymptotic
result is completely indistinguishable from the exact result, while for $\beta=0.1$
and $\beta=0.15$ they are almost indistinguishable. For $\beta>0.2$ the asymptotic
approximation becomes inaccurate.}
\label{asymptoticGFuncN}
\end{figure}

Figure~\protect\ref{asymptoticGFuncN} shows the asymptotic results calculated
with eq.~(\protect\ref{G5_bigN_4}) versus exact results according to eq.~(\protect\ref{G5}).

It is to be mentioned that the case of small $\beta$ and large $N$ analyzed
here fits the situations of currents in conducting materials or ion drift
currents, mentioned in the introduction. In a conducting loop, the number of
charges may be of order of $10^{23}$, and $\beta$ may be of order $10^{-12}$,
so that $\beta^{2N}$ results in completely unmeasurable radiated power. In a
ion drift device, the number of charges may be of order of $10^{10}$ and
$\beta$ may be of order $10^{-6}$, so that the radiated power is somehow
bigger than for the conducting loop, but still unmeasurable.

\section{Conclusions}

The purpose of this work was to learn how the radiation from discreet charges
vanishes in the continuum steady state limit. We used a canonical configuration
of charges in uniform circular motion, uniformly spread around a circle.

We found that the log of the power decreases almost linearly with the increase
in the number of charges, and arrived to a close form solution to calculate
this power if the number of charges is big - see
Figure~(\protect\ref{asymptoticGFuncN_infty}).

Specifically, for low velocities, we derived a simple expression for the radiated
power. It shows that the radiated power is governed by $\beta$
at the power of twice the number of charges - see eq.~(\protect\ref{G5_bigN_4})
and Figure~(\protect\ref{asymptoticGFuncN}), explaining why the radiation
in all the cases considered as DC is unmeasurable.

\end{document}